\documentclass[twocolumn,tighten,times,twocolappendix]{aastex631}

\usepackage{xspace}

\usepackage{comment}
\usepackage{amsmath}
\usepackage{CJKutf8}

\begin{document}


\title{Thermally inflated accretors in post-mass transfer binaries: Abell 35 and its class revisited}
\shorttitle{A35-type Systems and inflated accretors}
\shortauthors{Bhattacharjee et al.}

\correspondingauthor{Soumyadeep Bhattacharjee}
\email{sbhatta2@caltech.edu}

\author[0000-0003-2071-2956]{Soumyadeep Bhattacharjee}
\affiliation{Department of Astronomy, California Institute of Technology, 1216 E. California Blvd, Pasadena, CA, 91125, USA}

\author[0000-0002-6871-1752]{Kareem El-Badry}
\affiliation{Department of Astronomy, California Institute of Technology, 1216 E. California Blvd, Pasadena, CA, 91125, USA}

\author[0000-0002-4544-0750]{Jim Fuller}
\affiliation{TAPIR, Mailcode 350-17, California Institute of Technology, Pasadena, CA 91125}

\author[0000-0003-1247-9349]{Cheyanne Shariat}
\affiliation{Department of Astronomy, California Institute of Technology, 1216 E. California Blvd, Pasadena, CA, 91125, USA}

\author[0000-0001-6970-1014]{Natsuko Yamaguchi}
\affiliation{Department of Astronomy, California Institute of Technology, 1216 E. California Blvd, Pasadena, CA, 91125, USA}

\begin{abstract}

A small but growing class of binaries containing hot ($T_{\rm eff}\sim10^5\rm~K$) white dwarfs (WDs) and rapidly rotating, apparently subgiant companions -- including the prototype, Abell 35 -- show companions that are too large and luminous to be ordinary main-sequence stars yet too numerous to be explained as finely tuned near-twin binaries. We argue that these stars are instead main-sequence accretors temporarily inflated out of thermal equilibrium by recent mass transfer. For the subgiant of Abell 35, a new Gaia DR3 astrometric orbit ($P_{\rm orb} = 790$\,d) combined with updated photometric and spectroscopic constraints yield $T_{\rm eff} \approx 4925\pm75~\rm K$, $R \approx 3\pm0.05~R_{\odot}$, near-solar metallicity, and rapid rotation aligned with the orbit ($v_{\rm rot} \approx 195\pm3~\rm km~s^{-1}$), indicating substantial recent accretion and spin-up. Dynamical mass limits disfavor a coeval twin-binary origin, supporting the inflated-accretor interpretation. We test this scenario using self-consistent MESA binary evolution calculations with a new accretion prescription in which accreted material retains a fraction of its infall energy, against the default of setting its entropy to that of the accretor's surface. The accretor expands to giant-like radii when $\dot{M}$ is high yet remains within its Roche lobe, allowing stable mass transfer even for mass ratios traditionally considered unstable. After mass transfer ceases, the star contracts on Myr timescales through a bloated, rapidly rotating phase whose temperatures, radii, and spins match those observed in Abell 35–type systems. This framework naturally explains the population and unifies Abell 35-type binaries with post-AGB binaries, blue lurkers, and wide WD+main-sequence systems as successive stages of the same post-mass-transfer evolutionary pathway.

\end{abstract}


\section{Introduction}\label{sec:intro}

Mass transfer is a crucial phenomenon in binary systems which can dramatically alter the evolution of the stars \citep{Mathieu25}. In this work, we focus on low mass stars which form white dwarfs (WD). Interaction is most frequently triggered when the more massive star evolves off the main sequence to the giant branches and fills its Roche lobe. Stable mass transfer that begins when the primary is a subgiant (SG) or red giant branch (RGB) star forms hot subdwarfs (e.g. \citealt{Han03}), low-mass red giants (e.g. \citealt{El-Badry22,Jayasinghe22}), and helium white dwarfs (e.g. \citealt{Sun18,Li19}). Most binaries formed through this process have orbital periods $\gtrsim$$10^2$~days. However, mass transfer from evolved and convective donors often leads to instability and common envelope (CE) evolution \citep{Ivanova13}. This typically results in short-period ($\lesssim$$1$~day) orbits. Short-period WD + main sequences binaries (e.g. \citealt{Rebassa-Mansergas07,Parsons16}), cataclysmic variables (e.g. \citealt{Meyer79}), and close binaries found inside planetary nebulae (PNe, \citealt{Bond90Binary,Jones20}) are usually assumed to have formed via CE evolution.



There is, however, an emerging population of post-interaction binaries which appear to challenge this classical boundary between stable and unstable mass transfer. The binary separations of these systems fall between the predictions of models for CEE and stable mass transfer, with periods of $\sim$$100$ to $\sim$$1000$ days. A handful of such systems were serendipitously discovered in spectroscopic and photometric surveys over the last few decades, including IK~Pegasi \citep{Wonnacott93} and the Kepler self-lensing binaries \citep[SLBs;][]{Kruse14,Kawahara18,Masuda19,Yamaguchi24}, but a large population of such objects was discovered with astrometric orbits from the Gaia mission \citep{Shahaf23,Yamaguchi24a,Yamaguchi24b}. Several works have modeled these binaries as having formed via CEE with an asymptotic giant branch (AGB) donor and efficient envelope ejection (e.g. \citealt{Yamaguchi24b,Belloni24}). But recent studies have increasingly favored a stable mass transfer origin of these systems \citep{Ge20,Temmink23,Hallakoun24,Yamaguchi25}, appealing to models in which mass transfer from an AGB donor remains stable over a wider range of mass ratios than classically assumed (e.g. \citealt{Ge20,Temmink23}).


Mass transfer can have profound effects on the accreting star (henceforth, the accretor), especially when it is stable. The accreted material carries both angular momentum and energy. The former can spin up the accretor to near-critical rotation rates \citep{Packet81,Popham91}, such that anomalously rapid rotation in old stars is often a sign of recent accretion \citep[e.g.][]{Leiner19,Nine23}. 

The effects of the accretion energy on the accretor are uncertain but may be even more dramatic. Rapid accretion can temporarily throw the accretor out of thermal equilibrium, especially if the accretion timescale is shorter than the thermal timescale of the star. This is predicted to inflate the accretor to potentially orders of magnitude larger radii \citep{Kippenhahn77,Neo77,Fujimoto89,Lau24}. Thus, a main sequence accretor may attain the dimensions of a giant. This may have significant effects on the outcome of the binary interaction. For example, if the inflated accretor fills its own Roche lobe, it can lead to a CE. Thus, it is crucial to study this effect systematically. Observing systems in this phase of inflation is challenging, as the inflated state of the accretor is short lived: soon after the mass transfer stops, the accretor starts contracting back to the main sequence. Additionally, it is difficult to distinguish temporarily inflated accretors from stars that naturally expanded after they terminated their main-sequence evolution.

A natural place to search for inflated accretors is in post-stable mass transfer binaries hosting hot ($T_{\rm eff}$$\sim$$10^5$~K) young WDs, which must have just recently transferred some of their envelope to their companions. We thus introduce the class of systems which forms the focus of this work: Abell~35 (hereafter, A35) type systems, named after the prototype. The majority of the known systems are inside PNe. These are wide binaries with a hot WD. However, unlike the other systems, the companions are (sub)giants. The subgiants are fast rotating (telltale sign for a recent accretion history), with rotational periods in the range of $\sim$$1-10$~days. A concise account of these systems can be found in section~8 of \cite{Bond24}. The orbital period in only one of the eight systems\footnote{This includes WeSb~1, the central star of PN showing signatures of transiting debris disk \citep{Bhattacharjee25, Budaj25}. We introduce WeSb~1 as an A35-type system for the first time in this work, but the details will be presented in a separate letter.} have been known from radial velocity studies ($P_{\rm orb}$$\approx$$2700$~day in PN LoTr~5, \citealt{Jones17}). And, in this work, we report the $790$~day period of the prototype A35 obtained from Gaia DR3 astrometry solution, making it the second A35-type system with a known period. Three other systems (and also LoTr~5) have Gaia \texttt{RUWE}$>$$1.4$ which hints at similarly long periods. 

The highlight of these systems is the subgiant nature of the companion. It is possible that these are simply evolved stars. Given the young age of the WD, this can only be achieved with a twin binary (where the initial mass ratio is near unity). However, such an initial condition is too fine-tuned to explain the current and growing population of A35-type systems. Additionally, the lack of known main-sequence companions in similar binaries further renders this unlikely.

In this work, we explore the alternate possibility that the companions in the A35-type systems are not subgiants by evolution, but instead main sequence stars inflated by the recent accretion from the WD progenitor. We base our analysis on the protypical system, A35, which is the only system with a Gaia astrometric orbit, enabling estimation of the dynamical mass of the components. The rest of the paper is organized as follows. We present our photometric, astrometric and spectroscopic analyses of A35 in Section~\ref{sec:a35_phot_ast_spec}. Here we revise several of the stellar parameters inferred in previous works, and also demonstrate a potential tension between the dyanamical and evolutionary mass of the subgiant thus challenging an evolved subgiant scenario). In Section~\ref{sec:mesa_modeling} we use state of the art stellar evolution code named Modules for Experiments in Stellar Astrophysics (MESA, \citealt{Paxton11,Paxton15,Paxton19}) to construct detailed binary evolution models to explain A35-type systems as hosting inflated stars. We also relate these systems to other classes of post-mass transfer systems under a common evolutionary pathway. Finally in Section~\ref{sec:conclusions} we present our summary and conclusions.


\section{Abell 35}\label{sec:a35_phot_ast_spec}

\begin{deluxetable}{lcr}
\tablenum{1}
\tablecaption{Summary of the properties of A35}
\label{tab:a35_table}
\tabletypesize{\small}
\tablewidth{0pt}
\tablehead{
    \colhead{Parameter} &
    \colhead{} &
    \colhead{Value}
}
\startdata
&\emph{Gaia} Properties&\\
DR3 ID && 3499149202247569536 \\
RA (deg) && 193.38634 \\
Dec (deg) && -22.87299 \\
Parallax (mas)&&6.0339$\pm$0.0485 \\
$G$ && $9.3583$ \\
Distance (pc)&&165.73$\pm$1.33 \\
$G$ && $9.3583$ \\
$G_{\rm BP}-G_{\rm RP}$&&$1.1510$ \\
$\mu_{\alpha}$ (mas~yr$^{-1}$)&&$-61.611\pm0.048$ \\
$\mu_{\delta}$ (mas~yr$^{-1}$)&&$-13.875\pm 0.019$ \\
Radial Velocity && -- \\
Orbital Period $(P_{\rm orb})$ && $790.22\pm2.26$~d \\
Eccentricity $(e)$ && $0.0489\pm0.0064$ \\
Inclination $(i)$ && $26.09\pm0.83$ \\
\hline
&Stellar Properties&\\
Subgiant $T_{\rm eff}$&&$4925\pm75$~K\\
Subgiant radius&&$2.95\pm0.05~R_{\odot}$\\
Subgiant $\rm[M/H]$&&$0.00\pm0.15$\\
Subgiant $v_{\rm rot}\sin(i)$&&$86\pm5~\rm km~s^{-1}$\\
Subgiant $P_{\rm phot}$&&$0.767$~d\\
WD $T_{\rm eff}$$^{\#}$&&$80\pm10$~K\\
WD $\log(g)$$^{\#}$&&$7.2\pm0.3$~K\\
WD radius&&$\approx$$(1.7\pm0.2)\times10^{-2}~R_{\odot}$\\
\enddata
\tablecomments{$\#$ See Section~\ref{subsubsec:mass_of_wd} for a discrepancy between spectral and photometric properties.}
\end{deluxetable}

\subsection{Overview and previous works}

A35 is a nebula (top panel of Figure~\ref{fig:a35_nebula_a35_cmd_with_orbit}) which was first discovered by \cite{Abell55} and was classified as a PN. The central star (henceforth, we simply use A35 to denote the central stellar system rather than the nebula) was first studied in \cite{Jacoby81} who recognized the binarity of the system and apparently evolved nature of the companion (see Figure~\ref{fig:a35_nebula_a35_cmd_with_orbit}). The  temperature of the subgiant was estimated to be $\approx$$5000$~K and $\log(g)\approx3.5$ by photometric comparisons with standard stars. The fast rotation rate of the companion was identified in \citet{Thevenin97}, who estimated a projected surface rotation speed of $v\sin(i)$$\approx$$55~\rm km~s^{-1}$. They also reported Ba enhancement in the subgiant. The WD is hot and young with a $T_{\rm eff}$$\approx$$80$~kK and $\log(g)\approx7.5$ \citep[][implying a cooling age of $\sim$$2-5\times10^5$ years]{Herald02,Ziegler12}. The mass of the WD is highly uncertain with estimates (based on comparison with evolutionary models) spanning a broad range of $\approx$$0.5\pm0.07~M_{\odot}$ across different works \citep{Herald02,Ziegler12,Lobling20}. The separation between the two stars was estimated to be $18\pm5$~AU in \citep{Gatti98} using images obtained with the Planetary Camera on Hubble Space Telescope (HST).



There has been significant uncertainty in the distance to A35, which has often propagated to uncertainty in other parameters. \cite{Jacoby81} first arrived at a distance of $360$~pc by comparing the photometry of the subgiant with template stars. Later, the parallax measured from a Hipparcos single-star astrometric solution \citep{Perryman97} yielded a distance of $163^{+96}_{-58}$~pc, placing it significantly closer. \cite{Ziegler12} argued, however, that this distance is incompatible with the inferred WD parameters and preferred the larger subgiant-derived photometric distance.  

\subsubsection{New results from Gaia astrometry}
\begin{figure}[t]
\centering

\includegraphics[width=1\linewidth]{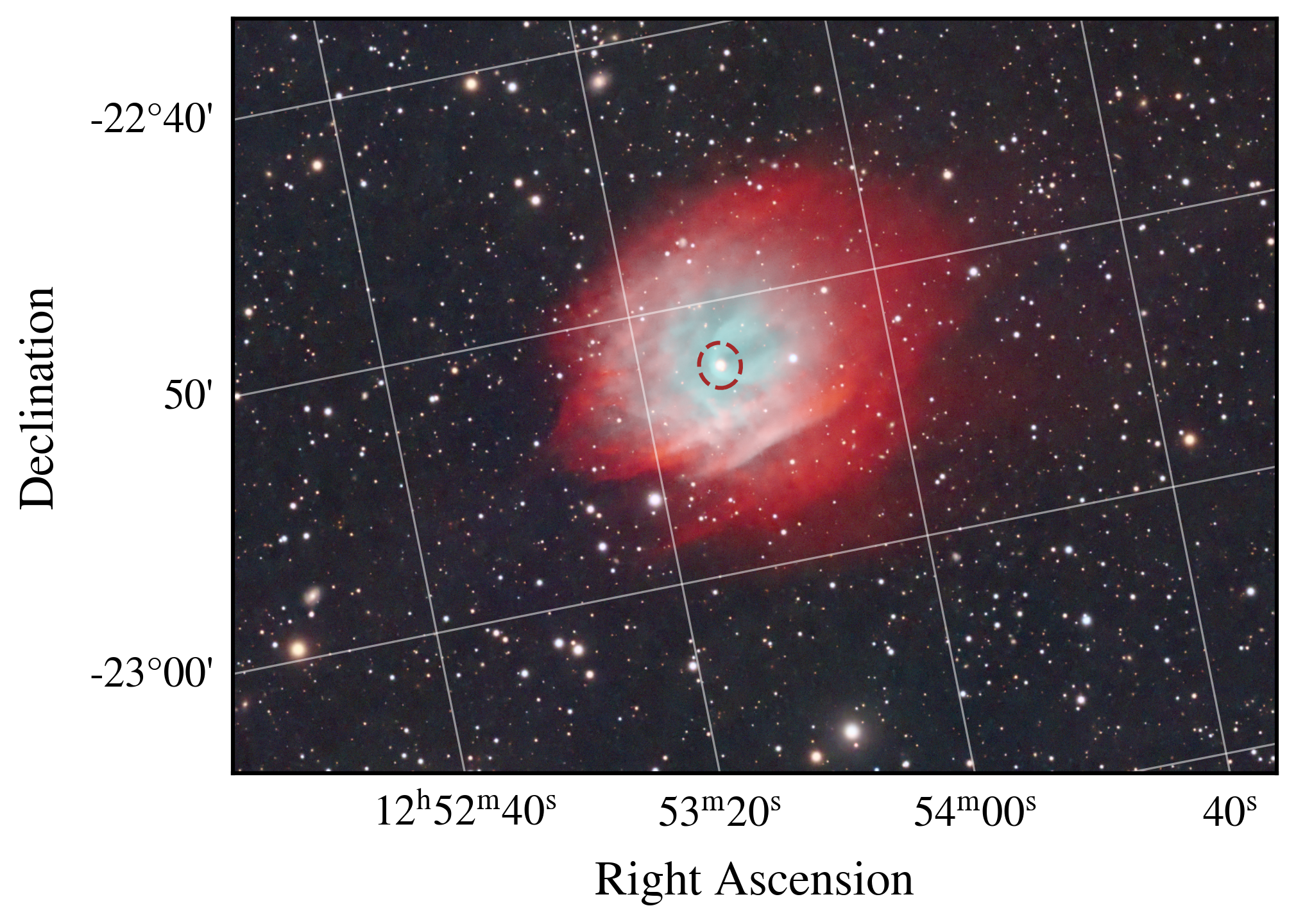}

\vspace{0.4cm}

\includegraphics[width=\linewidth]{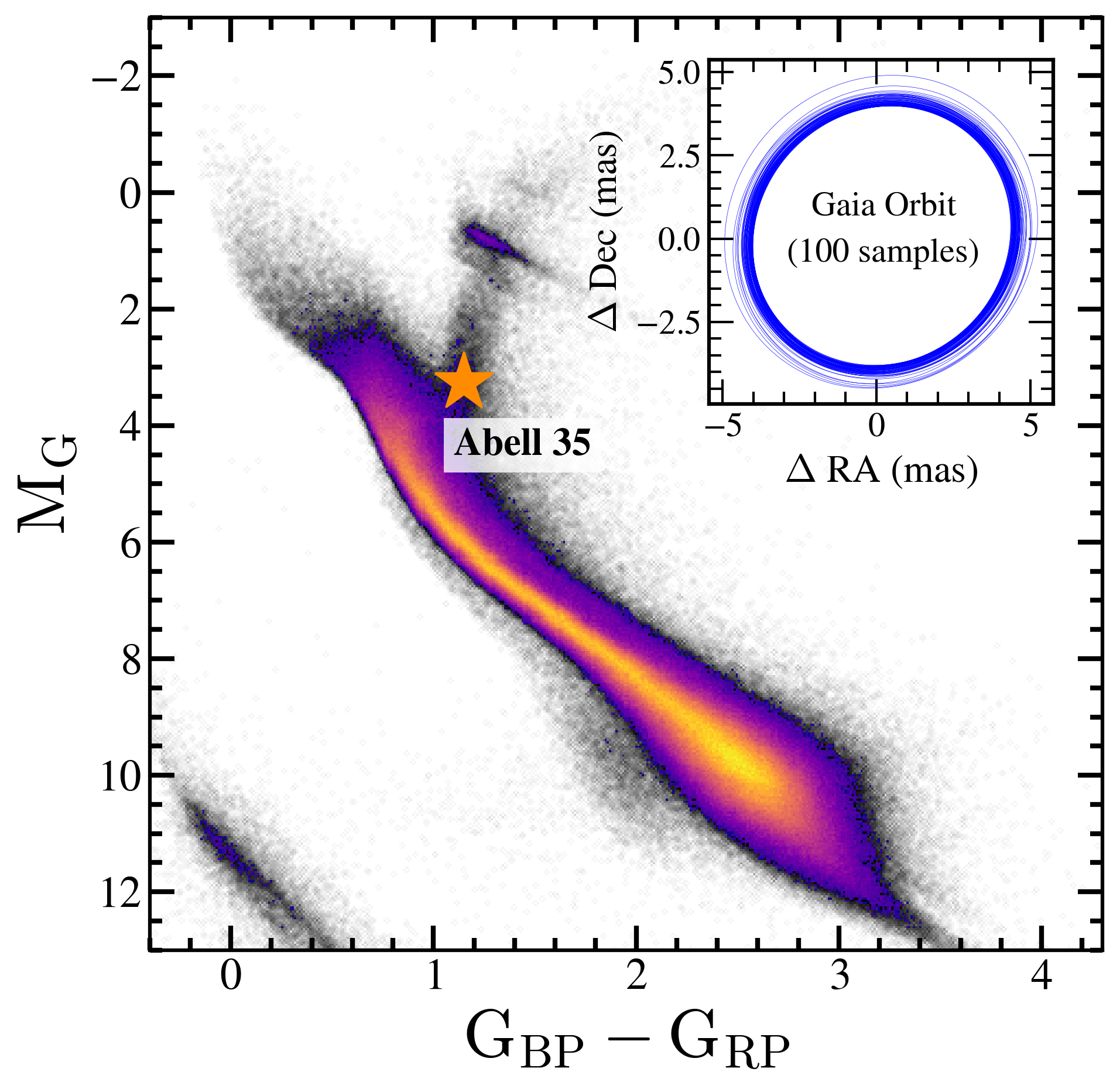}

\caption{Top panel: The nebula of Abell 35 (A35) with the central (unresolved binary) star marked. Henceforth, we simply call this central star as A35. Bottom panel: The Gaia color-magnitude diagram (CMD) showing the position of Abell 35 (effectively, the companion) in the subgiant branch. The inset presents 100 random samples of the Gaia orbital solution for visualization of the orbit.}
\label{fig:a35_nebula_a35_cmd_with_orbit}
\end{figure}

A35 has an astrometric orbital solution from Gaia DR3, whose parameters are listed in Table~\ref{tab:a35_table}. The Gaia parallax, corrected for the photo-center motion, is $6.034\pm 0.048$ (parallax error has been inflated as per prescription in \citealt{Nagarajan24})\footnote{Note that the correct parallax is the one provided with the astrometric solution. The single star parallax is higher at $7.30\pm0.32$ and quoted in \cite{Lobling20}. This is incorrect as A35 has a \texttt{RUWE} of 16 which makes the single star parallax unusable \citep{El-Badry25}.}. This yields a distance of $165.75^{+1.32}_{-1.30}$~pc, which is consistent with the Hipparcos measurement. The larger photometric distance favored by \cite{Jacoby81} and \cite{Ziegler12} is thus ruled out. Throughout the rest of this work, we use the Gaia distance for our analysis. The orbit has a period of $P_{\rm orb}=790\pm2$~days and is nearly circular with an eccentricity of $e=0.049$. Any reasonable estimates of the component masses yields a binary separation of $\lesssim$$2$~AU, much closer than estimated in \citealt{Gatti98}. The orbital inclination to the line of sight is low at $i=26.05$~degrees. Samples of the Gaia orbit is presented as an inset in the bottom panel of Figure~\ref{fig:a35_nebula_a35_cmd_with_orbit} for visualization.

We also discover that A35 is a part of a hierarchical triple system, where the post--mass-transfer inner binary is orbited by a distant tertiary companion identified at a projected separation of $9.56''$, corresponding to a physical separation of $\approx$$1600$~AU at the distance of A35. The tertiary (Gaia DR3 3499149202247569408) has a parallax ($6.013\pm0.030$~mas) and proper motion ($\mu_\alpha,\mu_\delta = -60.95\pm0.04,-13.39\pm0.02$~mas~yr$^{-1}$) that are statistically consistent with those of the inner binary (Table~\ref{tab:a35_table}), leading to a chance-alignment probability of $8.5\times10^{-5}$ \citep{Nagarajan24}. The Gaia photometry ($G_{\rm abs}=8.18$ and $G_{\rm BP}-G_{\rm RP}=1.94$) and the {\tt gspphot} stellar parameters ($T_{\rm eff}\approx4126$~K, $\log g\approx4.66$) indicate a late-type K dwarf with an estimated mass of $\approx0.65~{\rm M_\odot}$.\footnote{estimated using \url{https://www.pas.rochester.edu/~emamajek/EEM_dwarf_UBVIJHK_colors_Teff.txt}} Although the gravitational influence of this tertiary is negligible today, it may have been dynamically important earlier in the system’s evolution and played a role in bringing the inner binary to its current configuration (see Appendix~\ref{subsec:triple}). A detailed study of the role of the tertiary is beyond the scope of this work. But future statistics on the triple fraction of A35-type (or other wide-binary) systems would be informative.

\subsection{Stellar Parameters}

In this section, we derive constraints on some of the key stellar parameters of A35. A summary can be found in Table~\ref{tab:a35_table}.

\subsubsection{Spectral Energy Distribution}\label{subsubsec:sed_fit}

We use the spectral energy distribution (SED) to constrain the temperature of the subgiant, $T_{\rm eff,~SG}$, and the radii of both stars ($R_{\rm SG}$ and $R_{\rm WD}$). We use the synthetic SDSS griz, PanSTARRS y, and Generic/Johnson B and V photometry calculated from the Gaia XP spectrum \citep{Gaiasynphot}, and the Gaia $G$, $BP$, and $RP$ magnitudes. To incorporate the WD, we use the GALEX NUV and FUV magnitudes, and also the International Ultraviolet Explorer (IUE) HPDP photometry queried from \texttt{VOSA}. We adopt an uncertainty floor of $0.03$ mag in all bands to account for calibration uncertainties. The source was observed twice by GALEX, yielding significantly different FUV mags of $12.717\pm0.007$ mag and $12.332\pm0.005$ mag. In addition, the GALEX NUV photometry is $\approx$$0.1$ mag discrepant from the neighboring IUE photometry. This may indicate variability in the UV; we conservatively inflate the GALEX uncertainties to $0.2$ mag. We use the PHOENIX stellar models \citep{Husser13} for the subgiant and TMAP models \citep{Rauch03}\footnote{Downloaded from TMAP webpage: \url{http://astro.uni-tuebingen.de/~rauch/TMAF/flux_H+He.html}} for the WD to generate synthetic spectrum. We then use the \texttt{pyphot} package to calculate the magnitudes at the respective photometric bands. Finally, to perform the fit, we use \texttt{emcee} implementation of MCMC \citep{Foreman-Mackey13} with simple $\chi^2$ loss function.

For the subgiant, the two free parameters are $T_{\rm eff,~SG}$ and $R_{\rm SG}$. We keep the surface gravity fixed at $\log(g)=3.5$. This is motivated from the previous studies (and consistent with our final inferences). Additionally, the SED is not significantly sensitive to this parameter. We keep the metallicity fixed. We primarily use a solar  metallicity but also explore other values within the uncertainties derived in Section~\ref{subsubsec:hires_analysis}. For the WD, we rely on the parameter and abundance estimates in \cite{Herald02,Ziegler12}: $T_{\rm WD}=80\pm10$~kK, $\log(g)=7.5$ and $X({\rm He})=0.4$ and obtain the appropriate TMAP models. We use each model separately. Thus, the only free parameter for the WD is $R_{\rm WD}$. We use broad uniform priors for all our parameters. We keep the distance fixed at the Gaia distance of $165.75$~pc.


The uncertain parameter that significantly affects our results is the reddening. \cite{Herald02} estimated $E(b-v)=0.04\pm0.01$ using the Rayleigh-Jeans tail of the WD, whereas \cite{Ziegler12} suggests a lower range of $E(b-v)=0.02\pm0.02$. The reddening estimated from three-dimensional dust maps at the distance of A35, however, are higher: $E(b-v)\approx0.09$ mag as per \cite{Green19}\footnote{Queried using the online tool at \url{http://argonaut.skymaps.info/}} and $\approx$$0.06$ mag as per \citealt{Wang25}\footnote{Queried using the \texttt{Python} package \texttt{dustmaps3d}.}, with typical of $\lesssim$$0.015$ mag. 

\begin{figure}
    \centering
    \includegraphics[width=1\linewidth]{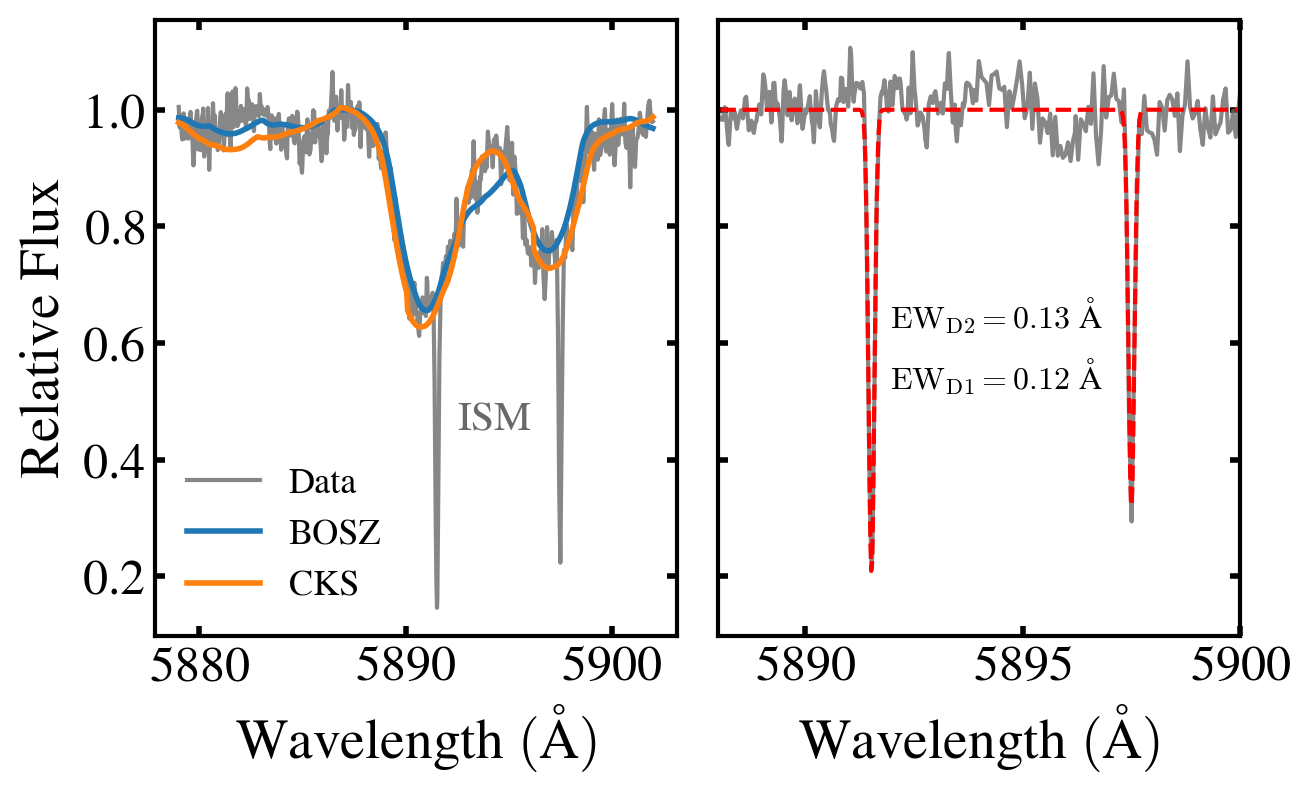}
    \caption{Inferring the reddening from the high resolution optical spectrum of the subgiant obtained with FEROS. Left panel: Normalized stellar spectrum around the Na-D doublet. The narrow ISM lines are well resolved. We overplot a (suitably rotationally broadened) synthetic spectrum from BOSZ and a stellar spectrum from the CKS survey best matching the observed Na-D lines in A35. The details about the spectral analysis are presented in Section~\ref{subsubsec:hires_analysis}. Right panel: Spectra further normalized by the best-matching templates. Gaussian profile fits to the ISM lines and the corresponding equivalent widths (EWs) are shown.}
    \label{fig:nad_reddening}
\end{figure}

We perform a fresh estimate of the reddening from the equivalent width (EW) of the interstellar Na-D doublet lines, resolved in our high resolution optical spectrosopic observations (presented in Section~\ref{subsubsec:hires_analysis}). We obtain a combined equivalent width of $\rm EW_{Na~D1+D2}=0.26$~\AA\ (see Figure~\ref{fig:nad_reddening}). Following the prescription in \cite{Poznanski12}, this yields a reddening of $E(b-v)\approx0.03$$\pm0.005$. Note that this is broadly consistent with the estimates from the WD spectra and photometry in \citep{Herald02,Ziegler12}, but significantly lower than those inferred from the dust maps. The reason behind such a discrepancy is unclear. However, \cite{Phillips13} notes that the error in \cite{Poznanski12} maybe underestimated. Their prescription inflates the error for Na-derived $E(b-v)$ to $0.02$, yielding a better agreement with the dust maps within error limits.

We perform the SED fit two ways. First, we keep $E(b-v)$\footnote{All dust extinction corrections were performed using the \texttt{dust\_extinction} \citep{dust_extinction, Gordon24_package_paper} implementation of the \cite{Karl23} galactic extinction models \citep{Decleir22, Gordon21, Fritzpatrick19, Gordon09}.} as a free parameter. The fit converges at $E(b-v)=0.047\pm0.023$ and yields a subgiant temperature of $T_{\rm rff,~SG}$$=$$4956\pm58$~K, radius of $R_{\rm SG}=2.93\pm0.03$ and a WD radius (with $T_{\rm eff,~WD}=80$~kK) of $R_{\rm WD}=(1.7\pm0.2)\times10^{-2}~R_{\odot}$. Manual inspection, however, shows that a $E(b-v)>0.04$ underestimates the WD flux in the far-UV when compared to the archival spectrum from the Far Ultraviolet Spectroscopic Explorer (FUSE). Thus, next, we fix $E(b-v)$ to the Na doublet-derived reddening of $0.03$. This yields a slightly lower subgiant temperature of $T_{\rm eff,~SG}$$=$$4914\pm19$~K, the same $R_{\rm SG}$, but a smaller WD radius of $R_{\rm WD}=(1.62\pm0.02)\times10^{-2}~R_{\odot}$. There is no appreciable change in the results with changing the metallicity within the limits discussed in Section~\ref{subsubsec:hires_analysis}. Thus, overall, after manual inspection of the dependency of the results on the various uncertain parameters, we provide the final range of subgiant parameters as: $T_{\rm eff}=4925\pm75$~K and $R_{\rm SG}=2.95\pm0.05~R_{\odot}$. Within the $T_{\rm WD}$ range of \cite{Ziegler12}, we find the WD radius to lie in the range of $R_{\rm WD}\approx(1.75\pm0.25)\times10^{-2}~R_{\odot}$.


\begin{figure}
    \centering
    \includegraphics[width=\linewidth]{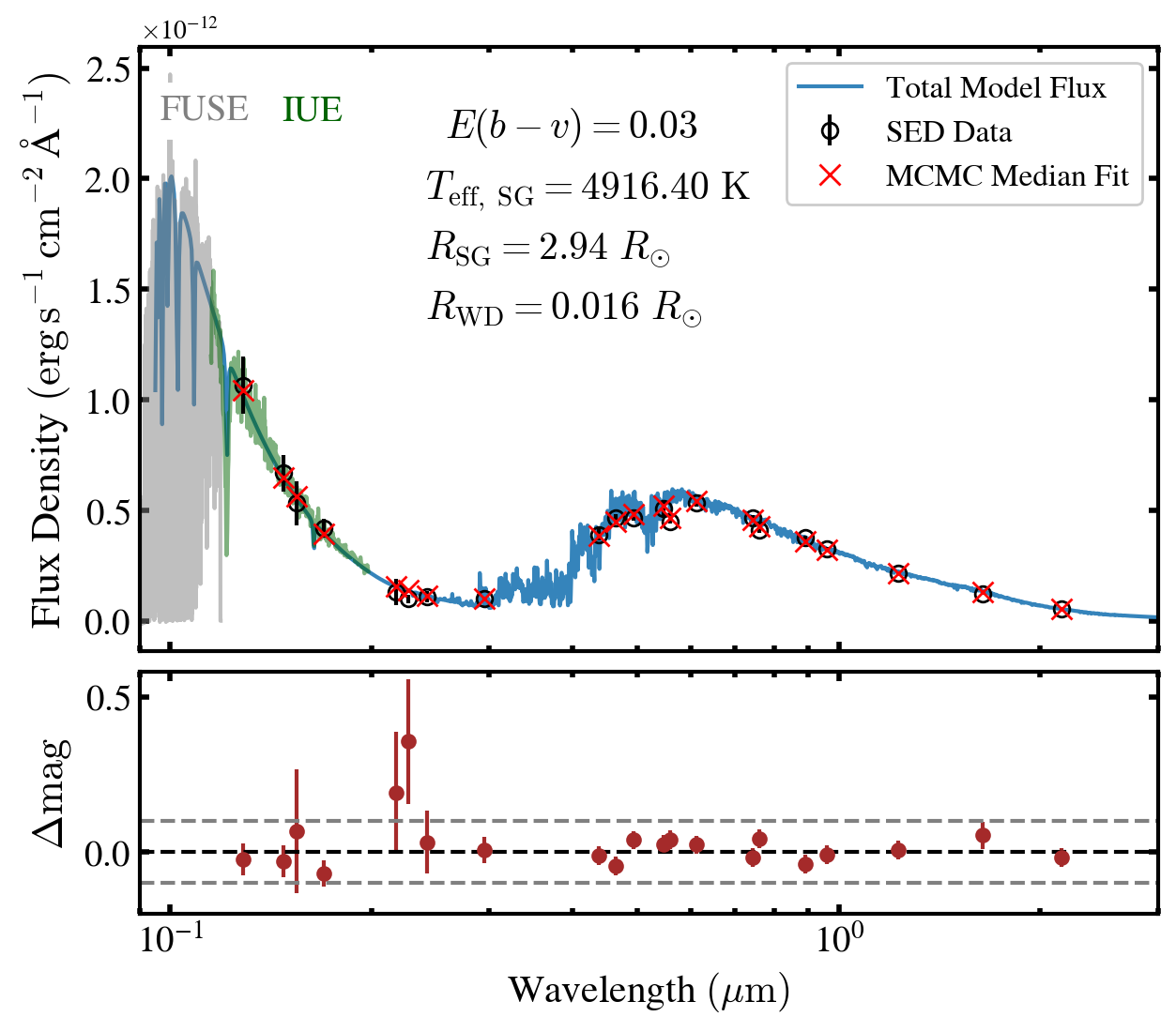}
    \caption{The SED of A35 (black circles) and our best fit PHOENIX+TMAP model with a fixed $E(b-v)$$=$$0.03$ (blue line). This fit uses a solar metallicity for the subgiant and a temperature of $80$~kK for the WD. The corresponding synthetic photometry from our model is shown as red crosses. We also present the FUSE (grey) and IUE (green) spectra. The PHOENIX spectrum is downresolved for clarity. The bottom panel shows the deviation of the best-fit synthetic magnitudes from those observed. The horizontal lines denote $\pm0.1$ mag derviation. The single discrepant point is GALEX NUV.}
    \label{fig:a35_sed_fit}
\end{figure}

\subsubsection{Spectroscopic analysis of the Subgiant}\label{subsubsec:hires_analysis}

We observed A35 with Fiber-fed Extended Range Optical Spectrograph (FEROS; \citealt{Kaufer99}) instrument installed on the MPG/ESO 2.2m telescope at the La Silla Observatory (program 115.28KE.001). We obtained six observations of $600$~seconds exposure between April and August, 2025. We used 1×1 binning. This allowed us to achieve a spectral resolution $R\sim50,000$ over $3860$\,\AA\ to $6770$\,\AA. For a single observation, the median signal-to-noise ratio (SNR) achieved was $\approx$$60$. We reduced the raw data using the CERES pipeline \citep{Brahm17}.

Prior to performing any analysis, we pseudo-continuum normalize the spectra as follows. First, we clip $C=20$\,\AA\ from both ends of each spectral order, where the noise becomes large. We then smooth the spectrum using a $W_s=1$\,\AA\ box kernel to remove noise. Following this, use a moving box of width $W_n=10$\,\AA\ and determine the $N=10^{\rm th}$ largest data point of the smoothed data within the window. We then linearly interpolate on these points to define the pseudo-continuum. The choice of the aforementioned parameters of this routine is arbitrary. But our results are not significantly sensitive to details in the continuum normalization, since we normalize models and data using the same procedure. We merge orders into a single spectrum, co-adding individual orders in overlap regions.


\begin{figure*}
    \centering
    \includegraphics[width=\linewidth]{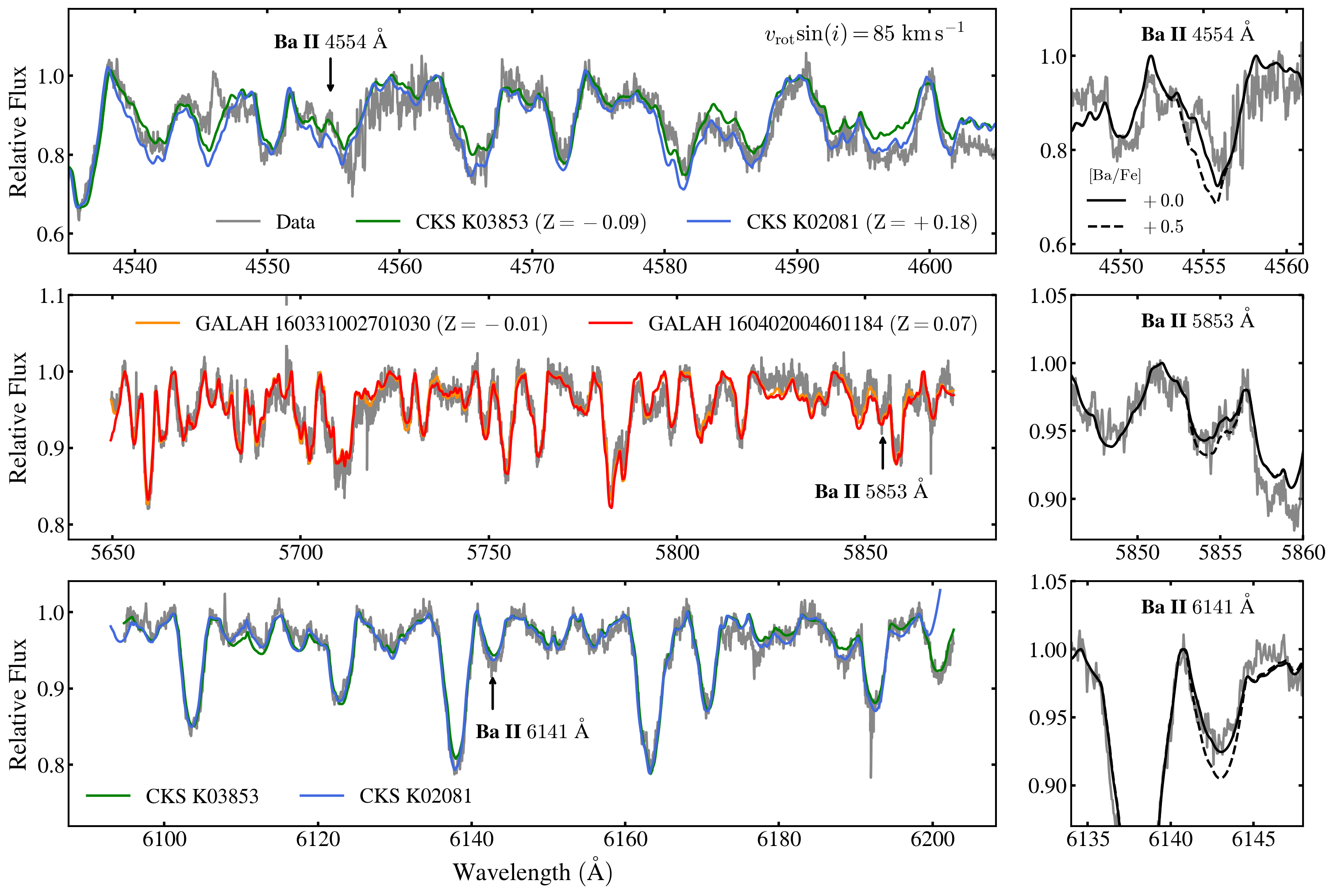}
    \caption{Left three panels: Comparison of the FEROS spectrum of the subgiant in A35 with GALAH and CKS spectra for three wavelength ranges, each containing a strong Ba~II line. The good match indicates a new-solar metallicity of the subgiant and no significant Ba excess. Right panel: Zoom-in on the Ba lines and comparison with synthetic spectra of $\rm[M/H]=-0.1$ from \texttt{iSpec}, with and without Ba excess. The latter yields a much better match indicating no significant Ba enhancement.}
    \label{fig:spectrum_comparison}
\end{figure*}

We do not detect any radial velocity (RV) variation in our spectrum. At the Gaia orbital period and inclination, any reasonable assumption of the binary component masses yields a RV semi-amplitude of a few kilometers per second. However, owing to the heavy rotational blending of the absorption lines, potentially augmented by surface magnetic wind/spot activity, our RV uncertainties are larger than this.\footnote{We see a scatter of $\approx$$5-10~\rm km\,s^{-1}$ in the measured RVs, depending on the models and spectral region used. But this is not a systematic error of the spectra as the ISM Na-D lines do not show this shift. Thus we attribute it to the overall RV uncertainty.} Thus, to improve SNR, we co-add all the six spectra. But we verify that our inferences do not change if the single epoch spectra are used instead.

The two primary properties we aim to constrain are the metallicity, $\rm [Fe/H]$ (where $\rm []$ is the standard notation for logarithm of the atomic number ratios with respect to the sun), and the projected surface rotational speed, $v_{\rm rot}\sin(i)$ of the subgiant. Accurately determining $\rm [Fe/H]$, in this case, is challenging due to the heavy rotational blending of the absorption lines, augmented by the uncertainties with spectral models. We thus experimented with several different approaches, which we discuss below. A combined estimate of the two parameters are listed in Table~\ref{tab:a35_table}.


\paragraph{Comprison with BOSZ models}
First, we used the BOSZ synthetic spectral library \citep{Bohlin17, Meszaros24}. We use templates with $R=50,000$, which is well-matched to the FEROS data. Based on the temperature constraint from the SED, we retrieved all spectra with $T_{\rm eff}=4750~\rm K$ and $5000~\rm K$ and $\log(g)=3.5$ from the publicly available database \footnote{\url{https://archive.stsci.edu/hlsp/bosz}}, and normalized them using the same routine described above. To avoid un-physical degeneracies, we restrict a few model parameters to values reasonably expected for a subgiant like A35. Specifically, we only consider models with alpha-element abundance with respect to total metal content of $\rm[A/M]\leq+0.25$ dex, carbon abundance of $\rm[C/M]=+0.0$ dex, and microturbulent velocity of $v_{\rm micro}\leq2~\rm km~s^{-1}$. This yields a total of $126$ templates. To incorporate the sensitivity of our results to the wavelength range being considered, we employ a heuristic method of dividing the FEROS spectrum into consecutive segments of $50$\,\AA. For each segment, we iterate through all the templates, finding the corresponding best-fit $v_{\rm rot}\sin(i)$ and RV over a suitable grid and record the $\chi^2$.\footnote{To estimate the wavelength-dependent error in the calculation of $\chi^2$, we first smooth the data using a Savitzky-Golay filter followed by computing the local standard deviation of the residuals from the smoothed data.} We estimate the best-fit parameters for each segment as the mean parameters of the 20 templates with the lowest $\chi^2$. Finally, we combine the estimates from all the segments by computing the mean and standard deviation. 


Overall, we obtain a net metallicity estimate of $\rm[M/H]$$\approx$$-0.04\pm0.2$. This is broadly consistent with that estimated in \cite{Thevenin97}($\rm[Fe/H]$$\approx$$-0.2\pm0.3$), and that estimated in \cite{Andrae23} from Gaia~XP spectrum ($\rm[M/H]$$=$$-0.14\pm0.15$)\footnote{This value has been obtained from R. Andrae over private communication, involving a revised calculation using the Gaia NSS parallax instead of the single star parallax used in the original work.}. A mild alpha enrichment of $\rm[A/M]$$\approx$$+0.05\pm0.2$, along with a microturbulent velocity of $\approx$$1.3\pm0.8~\rm km~\,s^{-1}$ are favored. We obtain a mean temperature of $\approx$$4917\pm118$~K, consistent with our inference from the SED. 

We obtain a more precise estimate of the projected rotational velocity of $v_{\rm rot}\sin(i)=86\pm5~{\rm km\,s^{-1}}$. This is significantly higher than estimated in \cite[][$\approx$$55~\rm km~s^{-1}$]{Thevenin97}. We now relate our revised estimate to the photometric period of $0.767$~days (noted in \citealt{Jacoby81}, and also recovered in the most recent TESS light curve\footnote{\url{https://www.tessextractor.app/}}, see \citealt{Bond24}). Assuming this to be the rotational period, the estimated radius of the subgiant gives an equitorial rotational velocity of $v_{\rm rot}=2\pi R_{\rm SG}/P_{\rm rot}\approx195\pm3~\rm km\,s^{-1}$. This implies an inclination of the rotation axis of $i\approx26.2\pm1.7$~degrees. This is in perfect agreement with the Gaia orbital inclination, and indicates an alignment of the rotational and orbital axes, as expected for accretion-induced spin-up.


\paragraph{Comparison with SPECTRUM models}
Next, we use another set of synthetic spectral models generated using the radiative transfer code SPECTRUM \footnote{https://www.appstate.edu/~grayro/spectrum/spectrum.html} \citet{Gray1994AJ}. We use the python interface provided by \texttt{iSpec} \citep{Blanco-Cuaresma2014A&A,Blanco-Cuaresma2019MNRAS}. We generate a grid of models varying $T_{\rm eff}$ between $4700$~K and $5400$~K at an interval of $50$~K, $\rm[M/H]$ between $-0.5$ and $+0.5$ at an interval of $0.05$, and $\rm[\alpha/Fe]$ between $0.0$ and $0.2$ at an interval of $0.1$. We keep surface gravity fixed at $\log(g)=3.5$. This yields a total of $900$ model spectra. \texttt{iSpec} generates normalized models, which we re-normalize using our custom method to maintain consistency with the data. We perform a similar analysis with the BOSZ models, where we chose the top $20$ best-fit models for a given segment and take the mean and standard deviation of the associated parameters.

Overall, we obtain similar results across different spectral regions, with the best-fit metallicity value lying in the range of $\rm[M/H]\approx-0.05\pm0.15$. This is consistent with the BOSZ estimate. These models, however, favor slightly lower temperatures, broadly in the range of $4850\pm75$~K, still consistent with the previous estimates. We find $\rm[\alpha/Fe]$ to mostly take values $\lesssim$$0.1$. The $v\sin(i)$ estimate remains consistent at $\approx$$86\pm2~\rm km\,s^{-1}$.

\paragraph{Comparison with GALAH spectra}
Spectral models do not perfectly reproduce observed spectra. Thus, we resort to another more empirical approach of comparing A35 spectra with other stellar spectra from high-resolution spectral libraries. We first use the GALAH DR4 HERMES spectral library \citep{Sheinis15,Buder25}. We use all spectra with $4700~{\rm K}\leq T_{\rm eff}\leq 5100~{\rm K}$, $3.25~{\rm K}\leq \log(g)\leq 3.7$ and ${\rm SNR}\geq 75$. This yields a total of $584$ spectra. The spectra are available over four channels ($B$, $V$, $R$, and $I$) spanning four different wavelength ranges. We use only the first three, as the fourth covers a wavelength range beyond our FEROS data. We again perform a $\chi^2$ minimization selection individually for the three channels. The third GALAH channel contains H$\alpha$, where A35 shows emission due to chromospheric activity and/or nebular contamination. We masked $\approx$$30$~\AA\ on either side of H$\alpha$ before calculating $\chi^2$. 

This empirical exercise favors higher metallicity values in the range of $\rm[Fe/H]$$\approx$$0.12\pm0.10$, still broadly consistent with the previous estimates at close to solar metallicity. The $v\sin(i)$ estimate from GALAH is consistent with the previous estimates of $85\pm3~\rm km~s^{-1}$. The $T_{\rm eff~SG}$ and $\log(g)$ estimates take values of $4950\pm60$~K and $3.4\pm0.1$, consistent with all previous estimates. A comparison of the A35 spectrum with two of the best-fit GALAH spectra in $V$-channel is shown in Figure~\ref{fig:spectrum_comparison}.

\paragraph{Comparison with CKS spectra}
To asses the robustness of our estimates, we use the Keck/HIRES spectra from the California-Kepler Survey (CKS, \citealt{Petigura17,Johnson17}), primarily to investigate the range of wavelengths not covered by the GALAH HERMES spectra. Because the CKS library is much smaller than GALAH, we only compare the FEROS spectrum of A35 to the CKS spectra of stars with parameters closest to the best-fit values from the GALAH comparison. We show the results in the first and third panels of Figure~\ref{fig:spectrum_comparison} for two CKS spectra of ${\rm [Fe/H]}=+0.17$ dex and ${\rm [Fe/H]}=-0.1$. dex\footnote{these are metallicity values estimated using spectra and isochrones jointly, \citealt{Johnson17} provided in \texttt{iso\_met} column in the CKS catalog} We find that the higher metallicity spectrum appears to fit some of the strong Fe line regions better, for example the line complexes around $4566$~\AA\ (first panel in the figure) and $6137$~\AA\ (last panel), which speaks against the lower metallicity inference from BOSZ (or previous studies). Overall, combining BOSZ, GALAH and CKS comparisons, we conclude that A35 is likely to be close to solar metallicity. 

\begin{figure*}[t]
    \centering
    \begin{minipage}{0.48\textwidth}
        \centering
        \includegraphics[width=\linewidth]{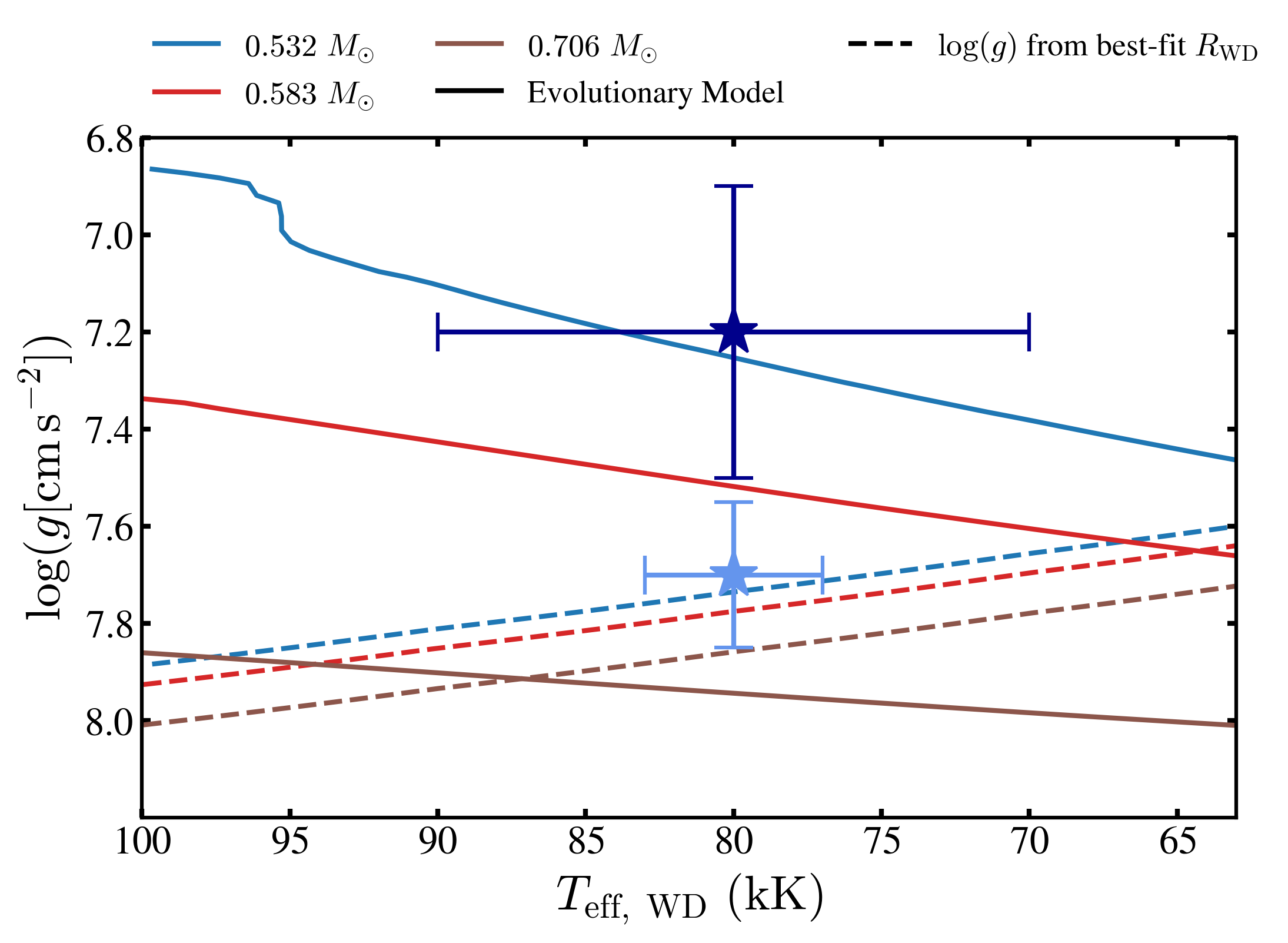}
    \end{minipage}\hfill
    \begin{minipage}{0.48\textwidth}
        \centering
        \includegraphics[width=\linewidth]{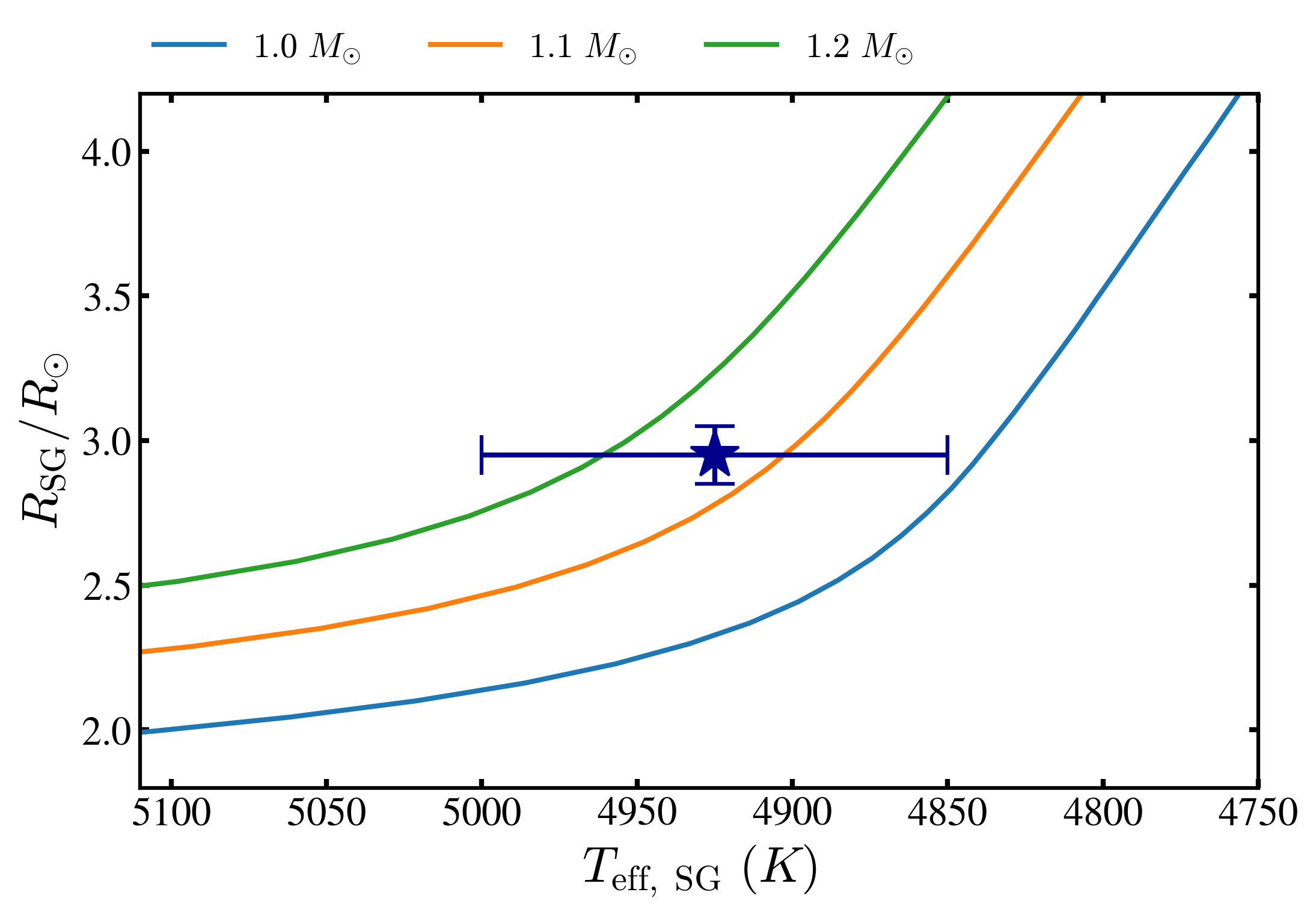}
    \end{minipage}

    \caption{Left panel: The spectrscopic $\log(g)$ and $T_{\rm eff}$ of the WD from \citet[][dark blue star]{Ziegler12} and \citet[][light blue star]{Herald02} compared to the post-AGB tracks from \cite[][solid lines]{Bertolami16} for $Z=0.01$ (model file \texttt{0100\_t03.dat.txt}). We also plot the $\log(g)$ derived from the radii that best fit the IUE photometry as discussed in Section~\ref{subsubsec:mass_of_wd} (dashed lines). Intersection of the solid and dashed lines within the error ellipse of A35 would signify a consistent solution. No consistent solution exists for the \cite{Ziegler12} parameters. A potential solution exists for $M_{\rm WD}$$\approx$$0.6~M_{\odot}$ with the parameter estimates from \cite{Herald02}. Right panel: The photometric parameters of the subgiant compared to the MIST models of solar metallicity, yielding a best evolutionary mass estimate of $1.1\pm0.1~M_{\odot}$. Uncertainty in metallicity results in an additional uncertainty of $\approx$$0.1~M_{\odot}$.}
    \label{fig:a35_millerb_compare_a35_mist_compare}
\end{figure*}

\paragraph{Lack of significant Ba enhancement}
Unlike \cite{Thevenin97}, we do not find evidence for significant Ba enhancement. In the left three panels of Figure~\ref{fig:spectrum_comparison} we mark three strong Ba lines at $4554$~\AA, $5853$~\AA, and $6141$~\AA. We find the GALAH/CKS spectra, none with any significant Ba excess, match the strength of the observed Ba lines quite well. We further investigate this by generating synthetic spectra using \texttt{iSpec} with varying Ba abundance. For this, we set $T_{\rm eff} = 4900\,$K, $\log(g)=3.4$, and conservative $\rm[M/H]=-0.1$. We present a zoom-in on the Ba lines in the right panels of the same figure, where we compare the A35 data with two \texttt{iSpec} spectra with $\rm [Ba/Fe]=0.0$ dex and $\rm[Ba/Fe]=+0.5$ dex. We find that the former fits the data significantly better than the latter. We note, however, that within the noise level of the FEROS spectrum, the Ba abundance cannot be obtained at a precision better than $\approx$$0.3$ dex. Thus any enhancement below this limit cannot be confidently ruled out. 



\subsubsection{Mass of the WD}\label{subsubsec:mass_of_wd}

The hot WD, BD$-$$22^{\circ}3467$, was first systematically studied in \citep{Herald02} using spectral data from IUE, HST, and FUSE. They arrived at a $T_{\rm eff}=80\pm3$~kK and $\log(g)\approx7.7\pm 0.15$. They reported a very broad range of WD mass of $0.5^{+0.5}_{-0.4}~M_{\odot}$. Later, \cite{Ziegler12} reanalyzed the data and revised the $\log(g)$ to a lower value of $7.2\pm0.3$. Comparing to the post-EHB tracks of \cite{Dorman93} and post-AGB tracks of \cite{Bertolami16}, they estimated a WD mass of $0.48\pm0.05~M_{\odot}$. However, the mass estimate was later revised in \cite{Lobling20} to a higher value of $0.533^{+0.04}_{-0.025}~M_{\odot}$. 

In the two latest studies, the distance to the source was assumed to be $\approx$$360$~pc. This yields a WD radius of $\approx$$0.03~R_{\odot}$, making it consistent with evolutionary parameters discussed in those works. However, as shown in Section~\ref{subsubsec:sed_fit}, the revised Gaia distance yields a much smaller radius. This leads to tension between the spectroscopic parameters and post-AGB evolutionary models. We demonstrate this in the left panel of Figure~\ref{fig:a35_millerb_compare_a35_mist_compare}. We plot the post-AGB tracks from \cite{Bertolami16} and the error margins of A35 from \citet[][dark blue]{Ziegler12}. To incorporate the constraints from the SED, we perform the following exercise. We take the $T_{\rm eff}$ values provided by the model and calculate a white dwarf radius that best-fits the IUE photometry. From the WD mass, we then calculate a predicted $\log(g)$\footnote{The TMAP models are available at a $T_{\rm eff}$ interval of $10$~kK. For intermediate temperatures, we performed a linear interpolation over models by keeping $\log(g)=7.5$ fixed.}. We call this the photometric $\log(g)$ and plot it in the same figure. We see that there is no consistent solution to the spectroscopic and photometric $\log(g)$ within the error margin provided in \cite{Ziegler12}. However, if we allow higher values of $\log(g)\gtrsim7.6$, a consistent solution is found at $M_{\rm WD}\gtrsim0.57~M_{\odot}$ and $T_{\rm eff}\lesssim75$~kK. In fact, a higher $\log(g)$ is consistent with the inferences of \citet[][shown in the same figure in light blue]{Herald02}, though their error in $T_{\rm eff}$ maybe underestimated. In fact, a consistent solution exists within the \citet{Herald02} parameter range for $M_{\rm WD}\approx0.62~M_{\odot}$.

The imperfect agreement between photometric, spectroscopic, and evolutionary parameters may imply that the uncertainties on the spectroscopic parameters are underestimated. It might also be a result of binary interaction leaving the WD with a thicker hydrogen shell than would be found for a single WD of the same mass.

\subsection{Disfavoring an evolved subgiant: A post-mass transfer inflated star instead?}\label{subsec:is_inflated}

As briefly discussed in Section~\ref{sec:intro}, the subgiant nature of the companion is intriguing. The key to distinguishing between a true evolved star from an inflated star is to look for discrepancy between the subgiant mass and the mass of the WD progenitor. This is because, given the young age of the WD, the former scenario requires a near-equal initial masses of the two components (twin binary; mass ratio $\lesssim$$1.015$ in our case). We first attempt to estimate the subgiant mass by comparing the SED-derived subgiant parameters to MIST \citep{Dotter16,Choi16} evolutionary models, with the evolved scenario being the null hypothesis. Within the metallicity uncertainty, this leads to a range of possible subgiant masses of $0.9~M_{\odot}$$\lesssim$$M_{\rm SG}^{\rm MIST}$$\lesssim$$1.3~M_{\odot}$. Assuming typical initial to final mass ratios (IFMR, \citealt{Cunningham24}), the WD's initial masses also fall in the this, thus appearing consistent. However, the uncertainties are too broad to detect any finer discrepancy. 



\begin{figure}
    \centering
    \includegraphics[width=1\linewidth]{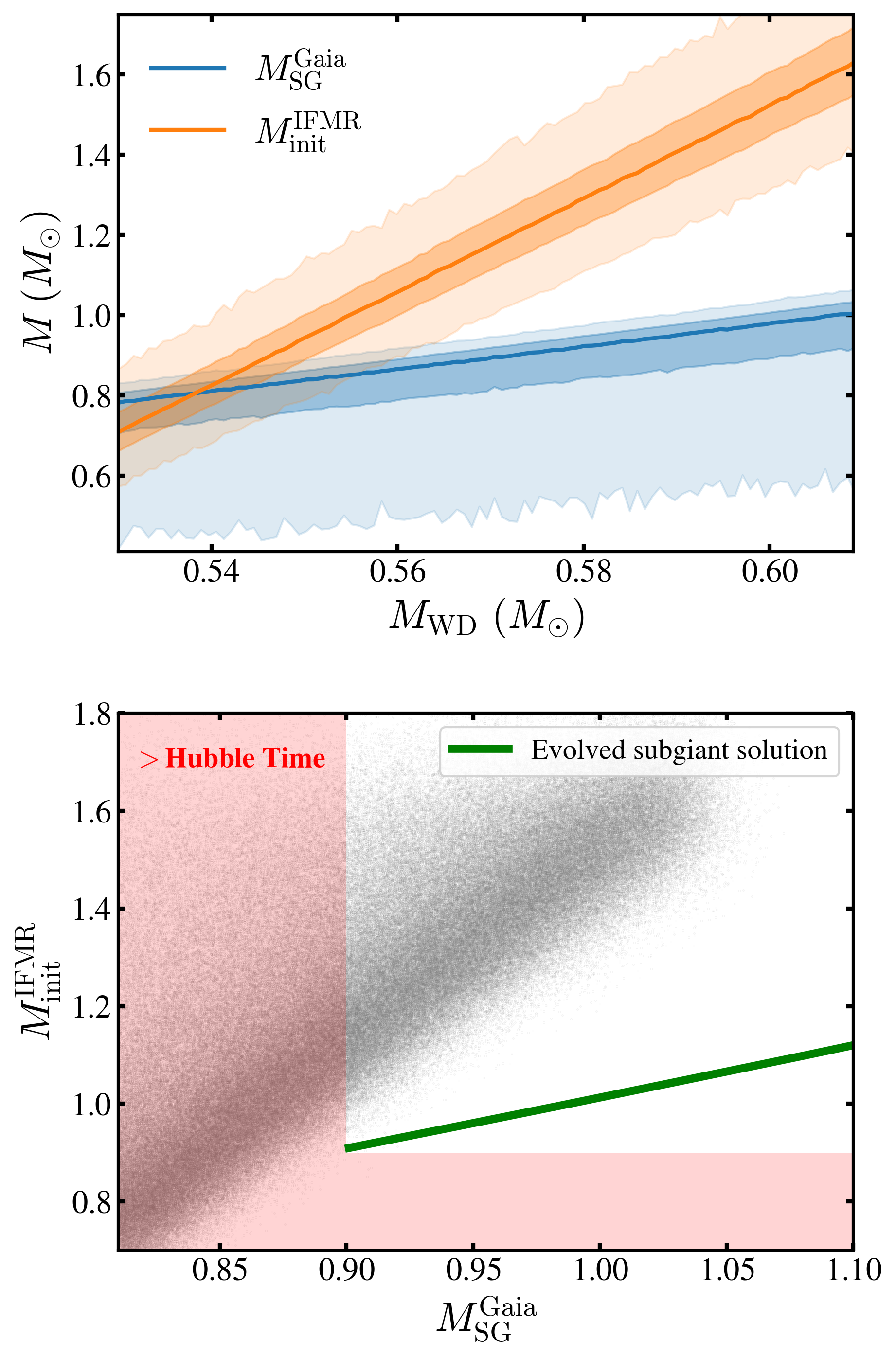}
    \caption{Examining the evolved subgiant scenario. Top Panel: Subgiant mass permitted by Gaia astrometry ($M_{\rm SG}^{\rm Gaia}$, blue) and the initial mass of the WD ($M_{\rm init}^{\rm IFMR}$, orange, estimated using the IFMR from \citealt{Cunningham24}), as a function of the assumed WD mass.  The dark (light) shaded region represents $(1\sigma)~3\sigma$-equivalent error region on either side of the median values (solid line). The intersection of the shaded regions is the twin-binary solution space. Bottom Panel: All the $M_{\rm SG}^{\rm Gaia}$ and $M_{\rm init}^{\rm IFMR}$ pairs from our uniform Monte Carlo sampling of WD masses. The region where either of the masses is $<$$0.9~M_{\odot}$ is shaded in red and excluded. The green line shows the MIST-estimated masses for which the companion can evolve to its present (subgiant) radius. A twin-binary solution would require the two components to fall on this line and thus is strongly disfavored.}
    \label{fig:minit_msbg_compare}
\end{figure}


The Gaia astrometric solution, however, provides an additional constraint for the componant masses. But, without any reliable RV information, the individual masses cannot be inferred uniquely. We thus perform the following exercise. We draw WD masses uniformly between $0.53~M_{\odot}$ and $0.61~M_{\odot}$. For each WD mass, we obtain the posterior of the Gaia astrometry-allowed subgiant masses. We do this by randomly drawing the Gaia orbital parameters within their respective error regions and solving the Keplerian orbital equation (see Appendix~\ref{app:comp_mass_kepler}). We employ appropriate correction for the WD flux contribution to the Gaia photocenter, with a G-band flux ratio of $F_{G,~\rm WD}/F_{G,~\rm SG}=0.9\%$. We then convert the assumed WD masses to an initial mass using the IFMR derived in \cite{Cunningham24}. This IFMR is not directly calibrated for initial masses $<$$1~M_{\odot}$. We perform a simple extrapolation of Equation~1 in their paper, which is adequate for our purpose. Finally, we compare the IFMR-derived initial masses with the astrometry-allowed subgiant masses. Given an initial mass, we use MIST models to estimate the mass of the companion which can evolve to its present (subgiant) radius at the appropriate age.

The result is shown in Figure~\ref{fig:minit_msbg_compare}. Firstly, very low subgiant mass of $\lesssim0.9~M_{\odot}$ can be ruled out since the main-sequence lifetime of such stars at the metallicity of A35 exceeds the Hubble time. For higher masses, it is evident that a twin binary solution is unlikely as the astrometry-allowed WD mass is much higher than can be produced by a star of the subgiant's mass. This discrepancy is further augmented by two additional reasons:


\begin{itemize}

    \item The IFMR used is for single stars. The binary interaction between the WD progenitor and the companion, however, has the potential to substantially truncate the mass of the resultant WD when compared to single stars of similar masses. This is witnessed in several recent works where the binary IFMR is seen to significantly deviate from that of single stars, especially for the lower end of WD masses (see, for example, \citealt{Ironi25, Shahaf25}). A truncation of the WD mass (i.e. higher $M_{\rm init}^{\rm IFMR}$ for a given $M_{\rm WD}$) will further disfavor a twin binary scenario.

    \item Assuming a fiducial $M_{\rm SG}\approx0.95~M_{\odot}$ yields a surface breakup speed of the star of $v_{\rm break}$$\approx$$250~\rm km\,s^{-1}$. This makes the measured $v_{\rm rot}$$\approx$$80\%$ of the break-up speed. This makes it reasonable to assume that, during mass transfer, the star was spun up to almost its break-up speed (post which it has spun-down to its present value). Assuming a Keplerian angular momentum of the accretion disk, the accretion mass needed to achieve the observed $v_{\rm rot}$ is given by:
    \begin{equation}
        M_{\rm acc}=\left(\frac{I_{\rm SG}}{M_{\rm SG}R_{\rm SG}^2}\right)\frac{v_{\rm rot}}{v_{\rm break}}M_{\rm SG},
    \end{equation}
    which, in this case, yields $M_{\rm acc}$$\approx$$0.06~M_{\odot}$ (assuming a fiducial sun-like radius of gyration). This needs to be subtracted from the current mass estimates to get the true initial mass of the subgiant. This would make the discrepancy with a twin-binary solution more pronounced. 

\end{itemize}
Finally, as already discussed in Section~\ref{sec:intro}, a twin binary origin of the whole population of A35-type systems is unlikely. This presumably extends to this object as well.

This shows that it is very unlikely for the subgiant to be a naturally evolved star. Rather, the hypothesis of it being a main-sequence star which is inflated from a recent mass transfer episode \citep{Neo77,Fujimoto89,Lau24,Zhao24} is favored. In the next section, we investigate this hypothesis further by exploring detailed binary evolution models with MESA.



\section{Accretor Response in MESA Binaries}\label{sec:mesa_modeling}

Several past works studying the response of the accretor have used single-star models accreting at a constant rate \citep{Kippenhahn77, Neo77,Fujimoto89,Zhao24,Lau24}. Though this has provided invaluable insights on the accretor response, a primary limitation of such an approach is its inconsistency with binary evolution models, which usually predict the mass transfer rate to vary by several orders of magnitude during the mass-transfer process. Such an approach also cannot self-consistently predict whether the mass transfer will proceed stably or the accretor will fill its own Roche lobe and trigger CEE. 

Here, we extend the previous works to full MESA binary simulations, evolving both the stars self-consistently through mass transfer. In our calculations, the accretor indeed inflates substantially, but remains within its Roche lobe, allowing the mass transfer to proceed stably. Post-mass transfer, the inflated accretor gradually relaxes back toward main sequence. We conjecture that the A35-type systems lie in this relaxation phase and present an opportunity to test models for accretor inflation. Additionally, we find that our models can potentially unify several post-mass transfer populations under the same evolutionary pathway.

\subsection{Method}
We use the latest MESA version of \texttt{r24.03.1}. We use the \texttt{MESABinary} module and set \texttt{evolve\_both\_stars=.true.}. We use the mass transfer prescription by \cite{Kolb90}. For the component inputs, we use a solar metallicity of $Z=0.014$ for both the stars. We use the \cite{Ledoux47} criterion to determine convective instability, use a mixing length \cite{Cox68} parameter of $\alpha=1.9$, and also include semiconvection with $\alpha_{\rm sc}=0.1$ \citep{Choi16, Temmink23}. We also model winds following \cite{Choi16} and \cite{Temmink23}: \cite{Reimers75} wind for RGB phase with a scaling factor of $\eta_R=0.1$ and \cite{Blocker95} wind for AGB phase with $\eta_B=0.2$. Additionally, we assume non-conservative mass transfer, with only $50\%$ of the transferred mass being accreted (i.e., $\beta = 0.5$; \citealt{Soberman97}; the rest being lost from the vicinity of the accretor).\footnote{Note that MESA defines $\beta$ as the fraction of the mass lost instead, so the \texttt{inlists} should be read accordingly.} We also include convective overshoot, similar to \cite{Yamaguchi24b} and \cite{Lau24}. We use a \texttt{step} overshoot scheme for the core, with \texttt{f/f0 = 0.129/0.0129} and \texttt{exponential} scheme for the shell with \texttt{f/f0=0.0174/0.00174}. We use \texttt{energy\_eqn\_option = 'dedt'}.

By default, we set \texttt{gold\_tolerances} to \texttt{.false.}. Our models run successfully through the mass transfer phases with \texttt{gold\_tolerances} turned on. But we do not show them because, for unclear reasons, they crash during the early WD cooling sequence and, thus, do not span enough for our figures that follow. We also tested that increasing the spatial\slash temporal resolutions do not affect the results significantly apart from increasing the run-time. For our fiducial A35 model (model~A, described in Section~\ref{subsubsec:model_overview}), we use here the output from a simulation with increased accretor spatial resolution of \texttt{mesh\_delta\_coeff=0.25} which yields results very similar to \texttt{gold2\_tolerance=.true.} but does not terminate prematurely.


\subsubsection{New Accretion Prescription}
The default accretion prescription of MESA is to set the thermodynamic properties of the accreted material (namely, the pressure, energy, and density) to those of the stellar photosphere. The rationale behind this treatment might be that most of the gravitational potential energy of the in-falling material gets radiated or dissipated away in the accretion disk, resulting in a smooth deposition of mass on the stellar surface. But this argument has certain limitations, as the accreted material may have much higher temperature and entropy than the stellar surface, because of the gravitational energy liberated during its in-fall. Whether it is shock-heated via direct impact accretion, or viscously heated in an accretion disk or boundary layer, the accreted material is expected to deposit a significant amount of excess energy\slash entropy.


On a different note, we also find that with the fiducial prescription, MESA fails to converge at high accretion rates of $\dot{M}$$\gtrsim$$10^{-3}~M_{\odot}\rm/yr$, especially for low accretor masses of $\lesssim$$1~M_{\odot}$. The reason is unclear, but it is likely of numerical origin and may be related to the steep entropy gradients at the surfaces of stars with convective envelopes.

We thus modify the thermodynamic properties of the accreted material via a custom accretion prescription in MESA. We do this by setting \texttt{use\_other\_accreting\_state=.true.} in the \texttt{inlist} for the accretor. We then use \texttt{run\_star\_extras.f90} to set the new prescription. We assume that the matter falls from infinity with no initial kinetic energy and is virialized upon reaching the stellar surface. We assume that only a fraction $f_E$ of the initial potential energy is retained and deposited onto the stellar surface. Considering a Keplerian accretion disk yields $f_E=0.5$. This yields the following prescription:
\begin{equation}\label{eq:acc_precription}
\begin{aligned}
v_a &= \sqrt{\frac{2f_EGM_{*}}{R_{*}}}; \\
E_a  &= (f_E - 1)\frac{v_a^2}{2f_E}; \\
\rho_a  &= \frac{\dot{M}}{4\pi R_{*}^2v_a}; \\
P_a  &= \rho_av_a^2, \\
\end{aligned}
\end{equation}
where $v_a$, $E_a$, $\rho_a$, and $P_a$ denote the radial velocity, total specific accretion energy, density, and pressure of the in-falling material, respectively. We assign $E_a$, $\rho_a$, and $P_a$ to the function arguments \texttt{total\_specific\_energy}, \texttt{accretion\_density}, and \texttt{accretion\_pressure}, respectively. For all our models we use $f_E=0.5$, but find that other values yield qualitatively similar results.\footnote{Inspection shows that not all values of $f_E$ are numerically stable, especially when $f_E\lesssim0.2$. Thus, we keep the study of the the quantitative dependence of our results on $f_E$ for a future more detailed work on accretion models.} Note that, in this new prescription, the density of the in-falling material is a function of $\dot{M}$, which is physically expected, unlike in the default MESA prescription which further justifies its revision. Comparison with the fiducial prescription is presented in Appendix~\ref{app:comparison_with_hamstars}.

\subsubsection{Estimating the accretor spin}
As discussed earlier, the accretor is expected to spin up almost to its breakup speed potentially after accreting only a small fraction of its mass \citep{Packet81}. The default settings of MESA are to enhance mass loss when the spin approaches breakup to prevent further accretion. However, several studies suggest that this may not be the correct picture, and the star can keep accreting matter (without gaining angular momentum, \citealt{Popham91}). Additionally, a model with rotation is more prone to numerical instabilities. Thus, we do not include rotation in the main analysis that follows. Instead, we calculate the stellar angular velocity, $\Omega$, as a post-processing step from the model outputs. Specifically, we employ a finite difference method using MESA's timesteps to compute $\Omega$ using the following evolution equation:
\begin{equation}\label{eq:omega_dot_de}
    \dot{\Omega} = \frac{\dot{J}_{\rm acc}+\dot{J}_{\rm MB}-\dot{I_*}\Omega}{I_*}.
\end{equation}
Here, $I_*$ is the moment of the inertia of the accretor which we consistently calculate inside our models as $I_* = \frac{2}{3} \int R^2dm$. $\dot{J}_{\rm acc}$ is the angular momentum gained through accretion, for which we assume the angular momentum of a Keplerian orbit at the stellar surface:
\begin{equation}
    \dot{J}_{\rm acc} = \dot{M}\sqrt{GM_{*}R_{*}}.
\end{equation}
$\dot{J}_{\rm MB}$ is the angular momentum lost through magnetic breaking, for which we use the prescription from \cite{Rappaport83}:
\begin{equation}\label{eq:mb}
    \dot{J}_{\rm MB} = -0.5\times10^{-28}\left(\frac{I_*}{M_*R_*^2}\right)M_*R_{\odot}^4\left(\frac{R_*}{R_{\odot}}\right)^{\gamma}\omega^3,
\end{equation}
and use $\gamma=4$ \citep{Rappaport83,Gossage23}. If at any timestep $\Omega>\Omega_{\rm crit}$, we force $\Omega = \Omega_{\rm crit}$. We recognize that this step may have non-trivial implications for a thermally contracting star. Conserving angular momentum, $\Omega\sim R^{-2}$ (assuming a roughly constant $I/MR^2$), whereas $\Omega_{\rm crit}\sim R^{-1.5}$. Thus, $\Omega/\Omega_{\rm crit}\sim R^{-0.5}$. Therefore, during contraction without any angular momentum loss, the rotation rate with respect to critical should, in fact, increase. However, by forcing $\Omega/\Omega_{\rm crit}=1$, we are artificially draining angular momentum. In reality, this may be driven by centrifugally enhanced mass loss from the star during this phase. Overall, we emphasize that this rotation rate calculation maybe far from reality. But Equation~\ref{eq:omega_dot_de} nevertheless provides a simple method to predict the star's rotation evolution that can be calibrated and compared to observations. 

In Appendix~\ref{app:mesa_with_rotation} we present some demonstrative MESA runs with accretor rotation and compare with our non-rotating models. But owing the added uncertainties with rotation in MESA, we do not include them in our main analysis. 

\subsection{Results}\label{subsec:mesa_results}

\begin{table*}[t]
\tablenum{2}
\centering
\caption{The other A35-type systems. For A35$^1$, refer Table~\ref{tab:a35_table}.}
\label{tab:all_a35s}
\resizebox{\linewidth}{!}{%
\begin{tabular}{ccccccccccccc}
Name & Gaia DR3 ID & RUWE & Distance & $T_{\rm eff,~SG}$ & $R_{\rm SG}$ & $T_{\rm eff,~WD}$ & $R_{\rm WD}$ & $P_{\rm orb}$ & $e$ & $E(b-v)$ & $P_{\rm phot}$ & $v_{\rm rot}\sin(i)$\\
 & & & (pc) & (K) & $(R_{\odot})$ & (kK) & $(0.01\,R_{\odot})$ & (d) &  &  & (d) & $\rm(km~s^{-1})$\\
\hline
LoTr 1$^{\rm 6}$ & 2917223705359238016 & $1.49$ & $1791.3^{+53.9}_{-47.7}$ & $4590$ & $9.7$ & $\gtrsim$$120$  & $\approx$$1.7$ &  &  & $0.07$ & 6.4 &  \\
WeSb 1$^{\rm 7, 9}$ & 423384961080344960 & $1.10$ & $3334.0^{+254.5}_{-310.2}$ & $6500$ & $6.0$ &  &  &  &  & $0.6$ &  & $90$ \\
LoTr 5$^{\rm 11, 12, 13}$ & 3958428334589607552 & $2.19$ & $449.7^{+8.9}_{-9.1}$  & $5100$ & $10$ &  &  & $\approx$$2700$ & $0.25$ & $0.0$ & 5.9 & $66$  \\
WeBo 1 & 465640807845756160 & $1.44$ & $1499.9^{+41.1}_{-49.5}$  & $4720$ & $7.5$ &  &  &  &  & 0.6 & 4.7 &  \\
Pa 27$^{\rm 2}$ & 1859955657931121536 & $1.07$ & $1436.6^{+22.7}_{-23.9}$ & 4760 & 8.55 &  &  &  &  & 0.28  & 7.6 &  \\
Hen 2-39$^{\rm 14}$ & 5256396485463285504 & $1.03$ & $8814.5^{+1929.3}_{-1614.2}$  & $4350$ & $14$ &  &  &  &  & $\approx$0.5 & 5.4 & $38^{+5}_{-5}$ \\
Abell 70$^{\rm 8}$ & 6907822573352460032 & $1.03$ & $3628.5^{+1459.4}_{-1010.8}$ & $\approx$$5000$ & $>$$1.5$ & $\gtrsim$$120$ &  & $\gtrsim$$720$ & $\gtrsim$$0.3$? & 0.07 & 2.07 & $37^{+5}_{-5}$ \\
UCAC2$^{\rm 10}$ & 1391041780057897856 & $2.51$ & $1200.9^{+39.4}_{-30.9}$ & $4790$ & $5.2$ & $105^{+5}_{-5}$ & $4^{+0.5}_{-0.4}$ &  &  & 0.0 & 1.99 & 81 \\
\end{tabular}
}
\tablerefs{\footnotesize 1)~This work, 2)~\cite{Bond24}, 3)~\cite{Herald02}, 4)~\cite{Ziegler12}, 5)~\cite{Lobling20}, 6)~\cite{Tyndall13}, 7)~\cite{Bhattacharjee25}, 8)~\cite{Jones22}, 9)~Bhattacharjee et al. in prep. 10)~\cite{Werner20}, 11)~\cite{Jones17}, 12)~\cite{Aller18}, 13)~\cite{VanWinckel}, 14)~\cite{Lobling19}}
\tablecomments{\footnotesize Except for Hen~2-39, where the distance is uncertain, we re-performed the SED fit, following the same method as Section~\ref{subsubsec:sed_fit}, to obtain the subgiant parameters consistently. All of our values are broadly in agreement to those previously reported in literature.}
\end{table*}

\begin{figure}
    \centering
    \includegraphics[width=\linewidth]{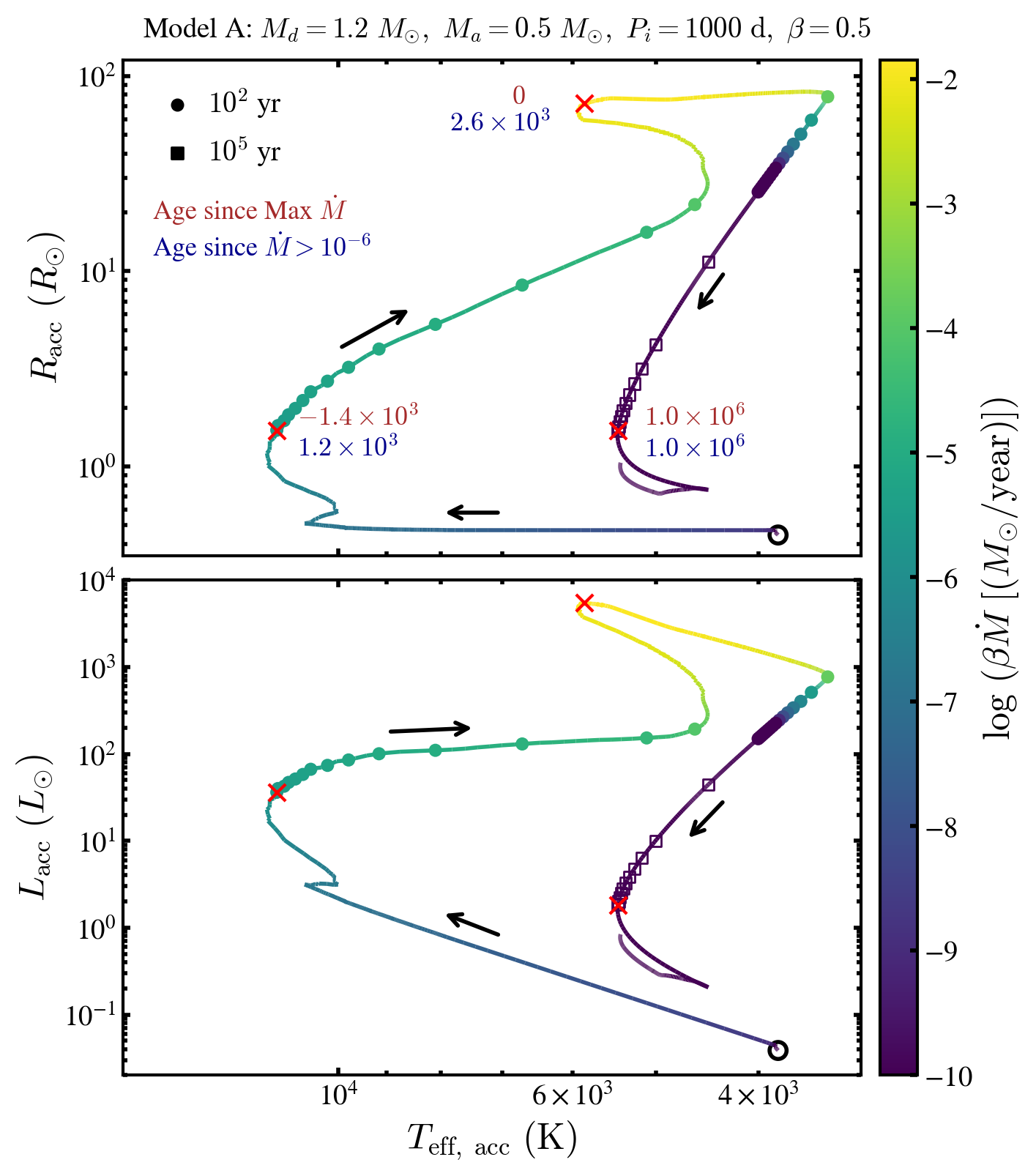}
    \caption{Top Panel: Evolution of radius ($R_{\rm acc}$) and temperature ($T_{\rm eff,~acc}$) of the accretor in model A over the course of the binary evolution, colored according to $\beta\dot{M}$. The black open circle marks the beginning of the model, and the arrows mark the direction of evolution. The filled circles (open squares) mark the intervals of $10^2$ ($10^5$) years -- only for the regime where $R_{\rm SG}>1.5~R_{\odot}$ to focus on the inflated state of the accretor. We also provide the ages, since beginning of mass transfer and since the maximum $\dot{M}$, at three representative epochs marked with red crosses. Bottom Panel: Evolution shown in the Hertzsprung-Russel diagram. Overall we find that the accretor rapidly inflates in $\sim$$10^4$ years and then begins to contract once mass transfer ends. The contraction slows as the star approaches the main sequence over $\sim$$10^6$ years.}
\label{fig:radius_vs_temp_a35_model_detail}
\end{figure}

\begin{figure}
    \centering
    \includegraphics[width=\linewidth]{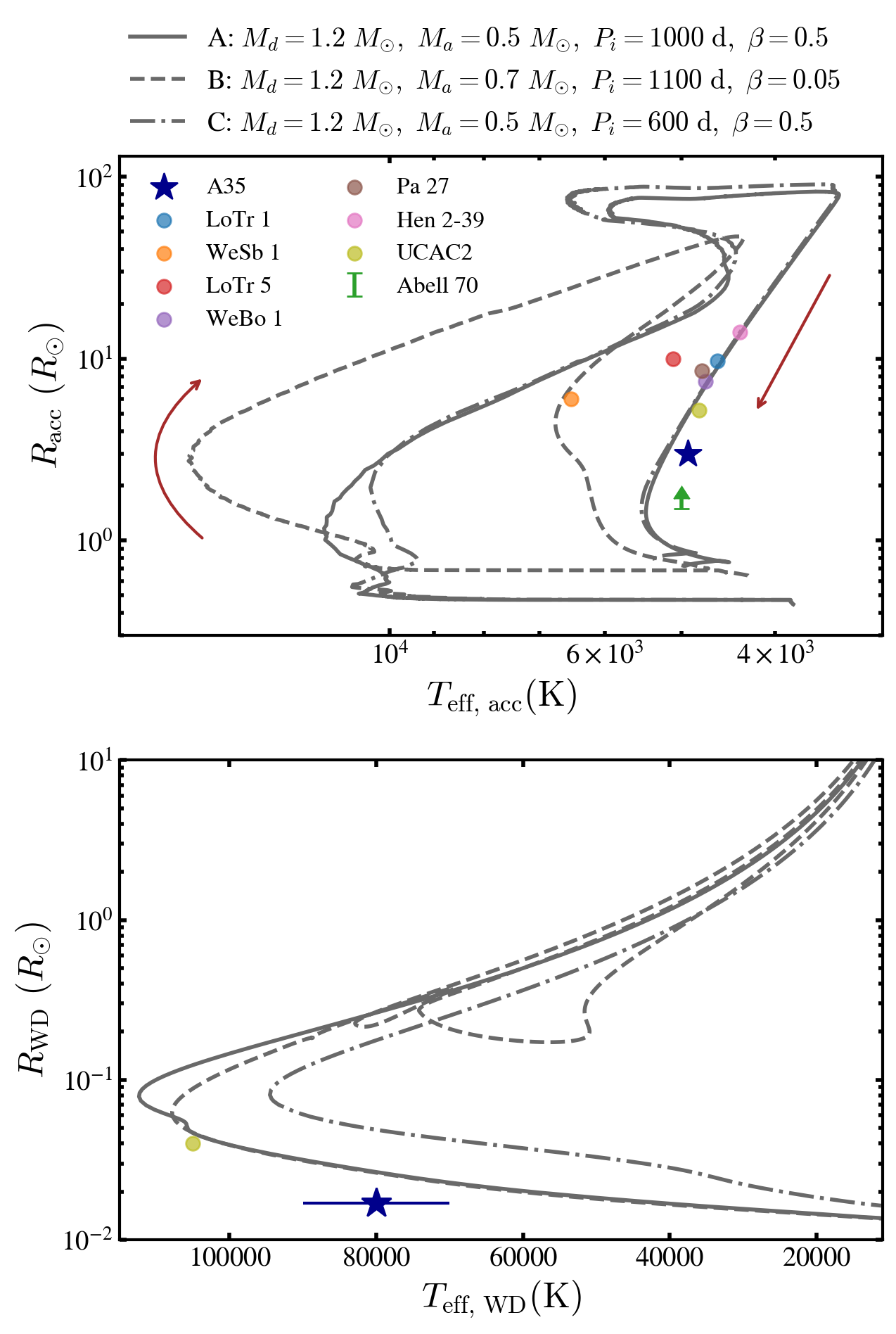}
    \caption{Top panel: Evolution of the accretor in $R-T_{\rm eff}$ space. We show three models with the same initial donor mass, but different accretor mass, initial orbital period, and $\beta$ (see label on top). We plot the stellar parameters of the subgiants in all known A35-type systems for comparison. Bottom panel: Evolution of the donor. We plot the WD parameters of A35 and UCAC2. We find the models to broadly reproduce the diversity of the observed population.}
    \label{fig:radius_vs_temp_a35}
\end{figure}


\subsubsection{Model overview}\label{subsubsec:model_overview}

We now present some models that approximately reproduce the observed properties of A35 and its class. We have not carried out an exhaustive search of the parameter space to best-match the system's properties but only endeavor to identify the typical evolutionary histories that can give rise to A35-type systems. We find that several initial configurations can reasonably produce A35-type systems, thus relaxing the need for any fine-tuning to explain this population. Overall, our models start with an unequal-mass binary ($q \sim 0.5$), where the more massive star ($M>1.0~M_{\rm \odot}$ for A35) evolves off the main sequence and transfers mass to its less massive companion ($M<1~M_{\rm \odot}$) with an initial period of $\sim$$10^3$~days. Depending on the exact parameters, the mass transfer can occur when the donor is on the RGB, forming a low-mass ($M_{\rm WD}\lesssim0.48~M_{\odot}$) WD, or when it is on the AGB, forming a more massive ($M_{\rm WD}\gtrsim0.53~M_{\odot}$) WD. The accretor gains a significant amount of mass, and expands in radius substantially, briefly reaching up to $\sim$$10^2~R_{\odot}$, simultaneously spinning up to critical rotation. It then contracts towards the main sequence, passing through an A35-like state around $\sim$$10^5$~years after mass transfer ends, when it is still inflated and rotating fast.

We will discuss three representative models (models A, B, and C) with different initial parameters. All the three models have an initial donor mass of $M_d = 1.2~M_{\odot}$. Models A and C have an initial companion (i.e. accretor) mass of $M_a = 0.5~M_{\odot}$ and $\beta = 0.5$ but differ in their initial period of $P_i=1000$~days and $600$~days, respectively. For Model B, we use a much lower $\beta=0.05$ with a period similar to Model $A$. For fair comparison, we make $M_{a,~B}=0.7~M_{\odot}$ so that the final mass of the accretor becomes comparable to the other two models.


We first discuss the overall response of the accretor and the change in its properties over the course of binary evolution and mass transfer. In Figure~\ref{fig:radius_vs_temp_a35_model_detail}, we present the trajectory of the accretor in $R_{\rm SG}-T_{\rm eff,~SG}$ space. For ease of discussion, we only present Model A, but the behavior is qualitatively similar for the other models (Figure~\ref{fig:radius_vs_temp_a35}). At the initial period of this model, appreciable mass transfer happens only during the AGB phase of the donor (Model C is different in this aspect as there the mass transfer occurs during RGB, forming a low mass WD), and driven by the thermal pulses. When the donor appreciably fills the Roche lobe, $\dot{M}$ rises steadily. We find that, across all scenarios, the accretor first becomes significantly hotter very quickly at the onset of accretion. This is followed by a rapid expansion and cooling over a timescale of $\sim$$10^3$ years. In this model, it reaches a peak radius of $\approx$$80~R_{\odot}$. The accretion rate attains very high values of $\sim$$10^{-2}~M_{\odot}\rm/year$ but only for a very brief period of $\approx$$20$~years. Soon after, $\dot{M}$ drops rapidly as the donor loses most of its envelope, leaving behind an inflated and spun-up accretor and a rapidly contracting donor. 


\subsubsection{Comparison with A35-type systems}\label{subsubsec:a35_compare}

The inflated accretor contracts over its thermal timescale, relaxing back to the main sequence, but still remaining appreciably inflated for around a million years. We propose that A35-type systems represent this phase of contraction where the accretor is still inflated and rotating fast. To show this, we compare the subgiant parameters of the known A35-type systems (see Table~\ref{tab:all_a35s}) to the MESA models. This is presented in the top panel of Figure~\ref{fig:radius_vs_temp_a35}. We clarify here that our models shown here are tailored for A35, and it is not expected that these will directly explain all the other systems. Nevertheless, we find that most of the systems lie close to our MESA model tracks. In the bottom panel, we also present the post-AGB evolution of the donor. Only two A35-type systems have spectroscopically derived WD parameters. We show them for comparison.

Focusing on A35, we find that model $A$ comes closest to explaining this system (both stellar and binary parameters), though we include the other two models for demonstrating the effects of the key model parameters. We note though, that none of our models can exactly meet the A35 parameters on all fronts. For example, in all our `best-fit' models, the subgiant temperature is $\approx250\pm50~\rm K$ higher than that of A35 at its radius. Such a discrepancy can arise from the following effects: 1) the spots on the stellar surface can bias its observed temperature to lower values, 2) the centrifugal force from the rotation and/or the magnetic pressure can compensate for thermal pressure, resulting in its lower temperature (see Appendix~\ref{app:mesa_with_rotation}). Given the complexity of the inflation process and significant uncertainties in several parameters and physics, we consider such differences acceptable.

\begin{figure}
    \centering
    \includegraphics[width=1\linewidth]{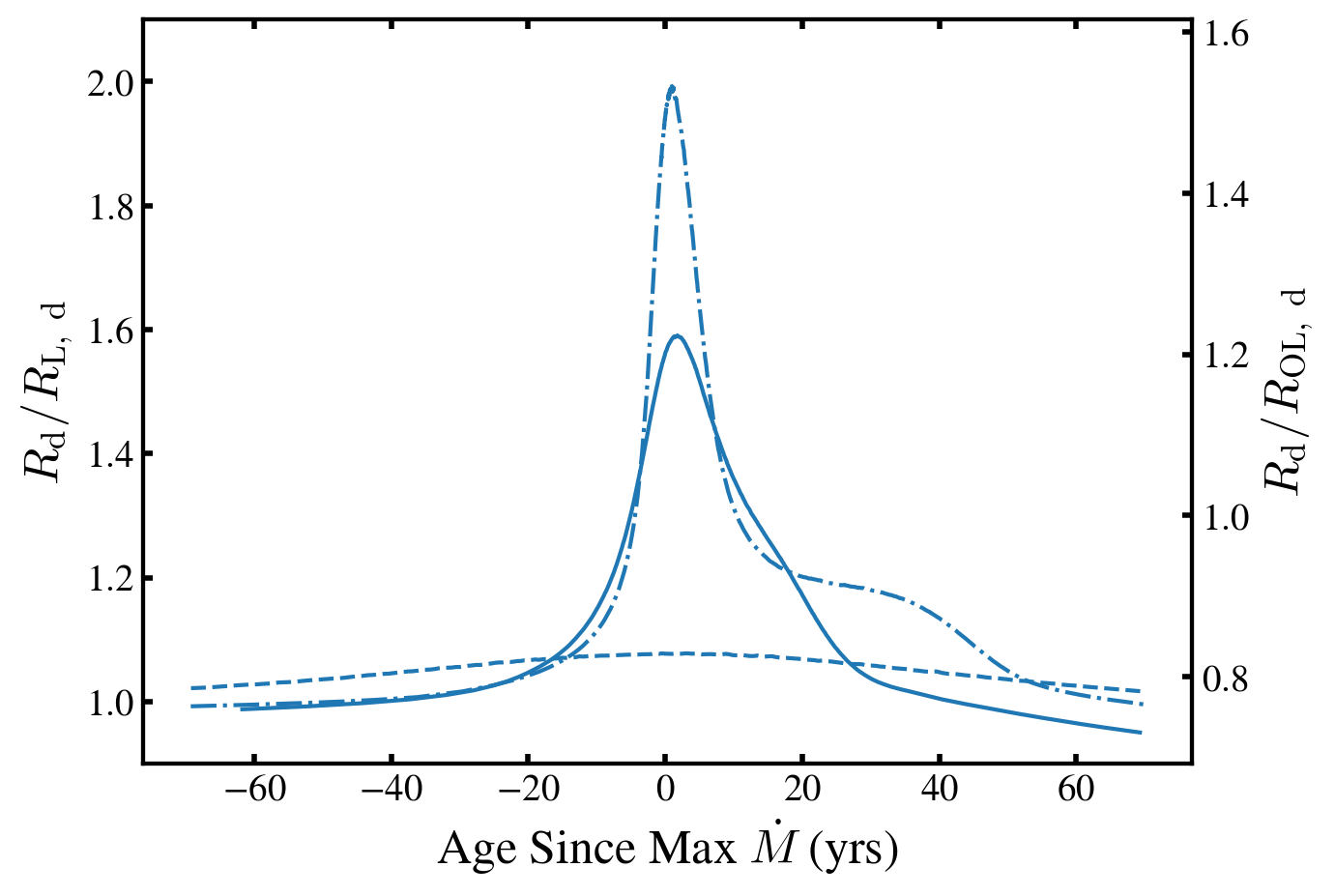}
    \caption{Roche Lobe ($R_{\rm L,\,d}$) and Outer lobe ($R_{\rm OL,\,d}$) overfill factors of the donor around the epoch of the highest accretion rates. For all our models, $R_{\rm OL,\,d}/R_{\rm L,\,d}\approx1.3$ during this epoch (calculated using equation~4 in \citealt{Temmink23}). Thus, we have used this single scaling value for the right axis. We follow the same linestyles for the models as in Figure~\ref{fig:radius_vs_temp_a35}.}
    \label{fig:stability_analysis_rlof_olof}
\end{figure}

\begin{figure}
    \centering
    \includegraphics[width=1\linewidth]{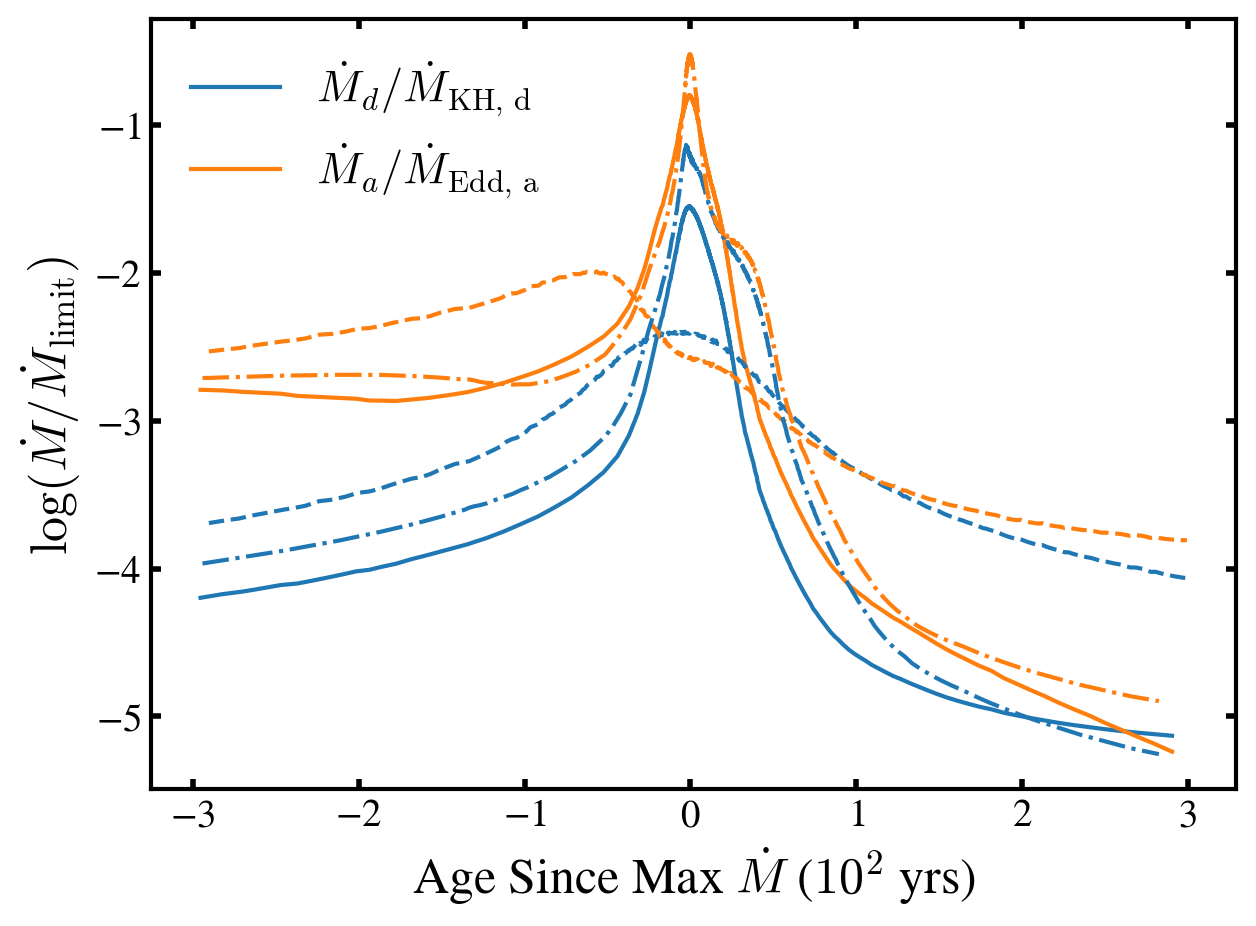}
    \caption{Ratio of the 1) donor mass loss rate rate, $\dot{M}_d\approx\dot{M}$, to the global thermal timescale of the donor $\dot{M}_{\rm KH,~d}$ (blue lines) and 2) accretor's accretion rate $\dot{M}_a\approx\beta\dot{M}$  to the Eddington-limited accretion rate $\dot{M}_{\rm Edd,~a}$. We only show $\pm$$300$~years around the epoch of peak mass transfer. Both the limits are always satisfied, indicating plausible stability. We follow the same linestyles for the models as in Figure~\ref{fig:radius_vs_temp_a35}.}
    \label{fig:stability_analysis}
\end{figure}

\subsubsection{Mass transfer stability at high $\dot{M}$}

A key uncertainty in our models stems from the very high mass transfer rates of $\gtrsim$$10^{-2}~M_{\odot}\rm/year$ that they reach during mass transfer that begins high on the RGB or AGB. Though the models do not fail to converge or experience accretor Roche lobe overflow, the evolution becomes uncertain at such high $\dot{M}$ values, potentially making them vulnerable to instability and CE. At the epochs of highest $\dot{M}$, the donor in models A and C achieve un-physically high values of Roche lobe overfill factors (albeit for a very brief period of a few tens of years, Figure~\ref{fig:stability_analysis_rlof_olof}, top panel) where the conventional mass transfer prescriptions break down. Additionally, the donor also overfills its outer (L3) lobe (same figure, only mildly for model A but substantially for model C). The associated mass loss from L3 can drain angular momentum, thus bringing the stars closer and triggering CEE. 

However, the study by \cite{Temmink23} suggests that neither of the above necessarily trigger an instability. Thus, we perform a few additional tests. We compare $\dot{M}$ to the accretion rate limited by the global thermal timescale of the donor (equation~6 in \citealt{Temmink23}):
\begin{equation}
    \dot{M}_{\rm KH,~d}=6.7\times10^{-7}\left(\frac{M_{d}}{M_{\odot}}\right)^{-1}\frac{R_d}{R_{\odot}}\frac{L_d}{L_{\odot}}M_{\odot}\rm/year.
\end{equation}
The mass transfer is likely to remain stable when $\dot{M}<\dot{M}_{\rm KH}$ \citep{Ge20}. Note that this is a stricter stability criterion than the quasi-adiabatic criterion derived in \cite{Temmink23}. Additionally, we compare $\beta\dot{M}$ (the effective accretion rate) to the Eddington rate of the accretor (equation~9 in \citealt{1977PASJ...29..249N}):
\begin{equation}
    \dot{M}_{\rm Edd,~a}=1.24\times10^{-3}\frac{R_a}{R_{\odot}}M_{\odot}\rm/year.
\end{equation}
We present this comparison in Figure~\ref{fig:stability_analysis} for all the three models. We find that our mass transfer rates are always contained within the two limits, showing that stability can possibly be maintained even at such accretion rates. This is in line with several recent works showing that mass transfer from RGB/AGB stars may frequently remain stable \citep{Ge20,Temmink23,Yamaguchi25}. Our models also show that although accretors expand significantly in response to mass transfer, they may often not fill their Roche lobes in AU-scale orbits (even with rotation, see Appendix~\ref{app:mesa_with_rotation}). Nevertheless, there may be additional factors driving instability which we discuss in Section~\ref{subsec:caveats}. We note here that in Model B, though $\dot{M}$ attains similarly high values, $\beta\dot{M}$ remains below $10^{-3}~M_{\odot}\rm/year$. The accretor thus gains very little mass. Nevertheless, it still inflates significantly showing that the accretor can inflate even without accreting a large amount of~mass.

\subsubsection{Time Evolution}

\begin{figure}
    \centering
    \includegraphics[width=\linewidth]{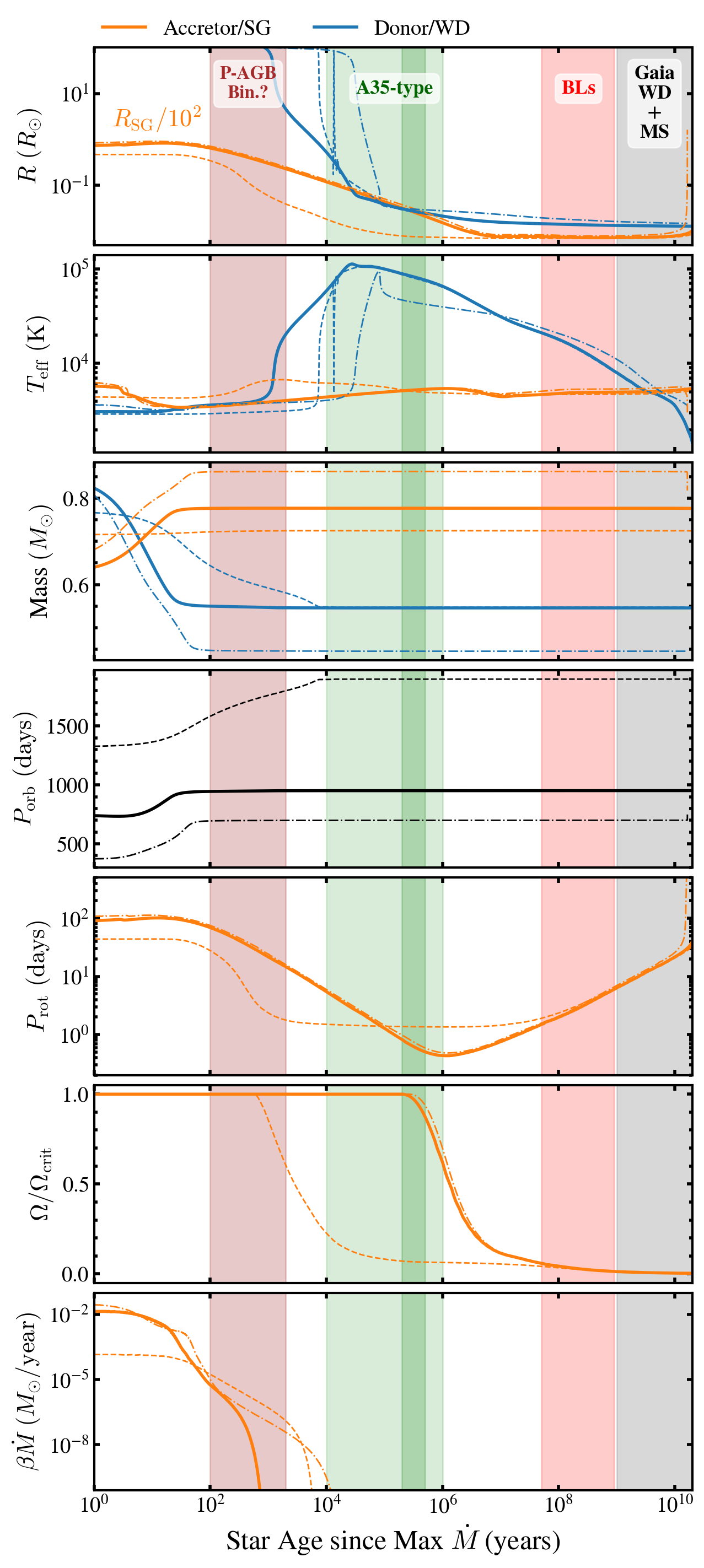}
    \caption{Evolution of stellar and binary parameters after maximum $\dot{M}$ for all three models (highlighting Model A for clarity). We denote the accretor and donor with blue and orange colors, respectively. The age range approximately satisfying the properties of the A35-type systems is shaded in green, with darker shading corresponding to A35 in particular. The age range for Post-AGB Binaries (P-AGB Bin.), Blue Lurkers (BLs) and Gaia WD+MS population are shaded in brown, red and gray, respectively. We follow the same linestyles for the models as in Figure~\ref{fig:radius_vs_temp_a35}.}
    \label{fig:params_vs_age}
\end{figure}

We now discuss how the stellar and binary properties evolve with time after mass transfer stops. We show this in Figure~\ref{fig:params_vs_age}. We concentrate on the periods after maximum $\dot{M}$. We find that the accretion rate decreases to $\sim 10^{-5} M_{\odot}\rm/yr$ within $100$ yr, and to $\sim 10^{-10} M_{\odot}\rm/yr$ within $10^4$ yr. By this time, both the stars have settled at their final masses. In Models A and C, the accretor gains a substantial amount of mass of $0.25~M_{\odot}$ and $0.35~M_{\odot}$, respectively. In Model B, due to low $\beta$, only a few $\times 10^{-2} M_{\odot}$ is gained. The binary period has also attained its final value of $\sim$$10^3$~days. During this time, the inflated accretor is relaxing back to the main sequence and the donor is rapidly contracting and heating up, forming a WD. We find that a post--mass transfer age range of $10^4-10^6$ years best correspond to the observed properties of A35-type systems (shaded in green in the figure). After a few million years, the accretor has relaxed back to the main sequence and the WD has entered the cooling~track.


It is worth discussing the evolution of the rotational speed of the accretor separately. As mentioned earlier, the accretor attains a critical rotation rate soon after the accretion of a small fraction of its mass, and remains critically rotating until the end of interaction. Critical rotation is also maintained during the initial contraction phase till the star has contracted to a radius $\lesssim$$10~R_{\odot}$. This is because, though the critical angular velocity ($\Omega_{\rm crit}$) increases as the star contracts, the rotation rate also increases to conserve angular momentum. This also leads to a decrease in $P_{\rm rot}$. Within the age range of A35-type systems, our models yield $P_{\rm rot}$$\sim$$1-10$~days, in good agreement with observations. At late times, magnetic braking takes over, draining angular momentum and increasing $P_{\rm rot}$ to $\mathcal O\rm(10~d)$ over the next $10^9$ years.


\subsection{Evolutionary connection to other populations}

\begin{figure}
    \centering
    \includegraphics[width=1\linewidth]{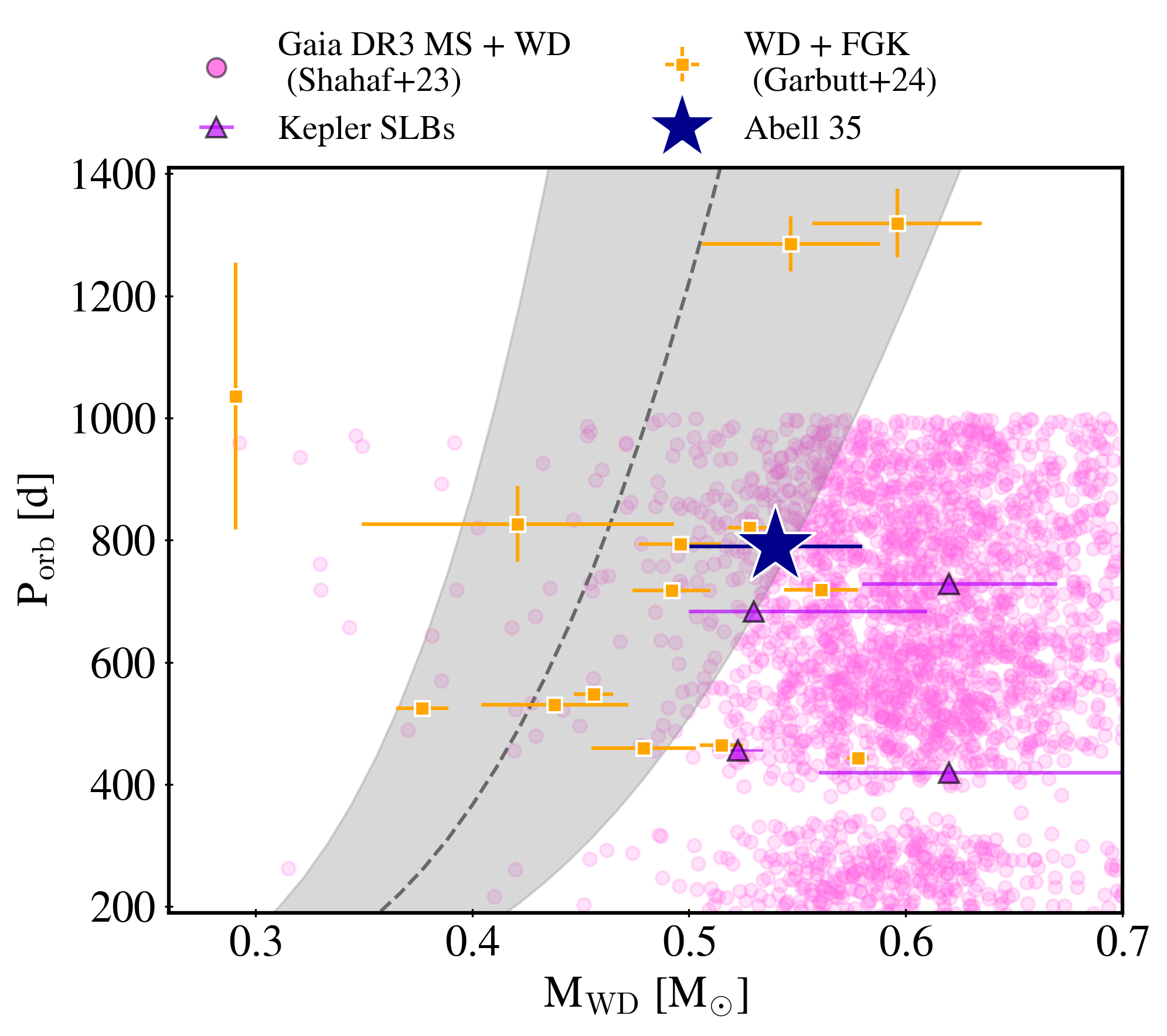}
    \caption{Orbital periods and WD masses for A35 and several other classes of related post-mass transfer binaries. The \cite{Rappaport95} relation for stable RGB mass transfer is shown as dashed line and a shaded error region. A35 seems most similar to the Gaia WD+MS binaries and the Kepler SLBs, which are both plausibly the results of AGB mass transfer.}
    \label{fig:porb_mwd_rappaport}
\end{figure}

\begin{figure*}
    \centering
    \includegraphics[width=1\linewidth]{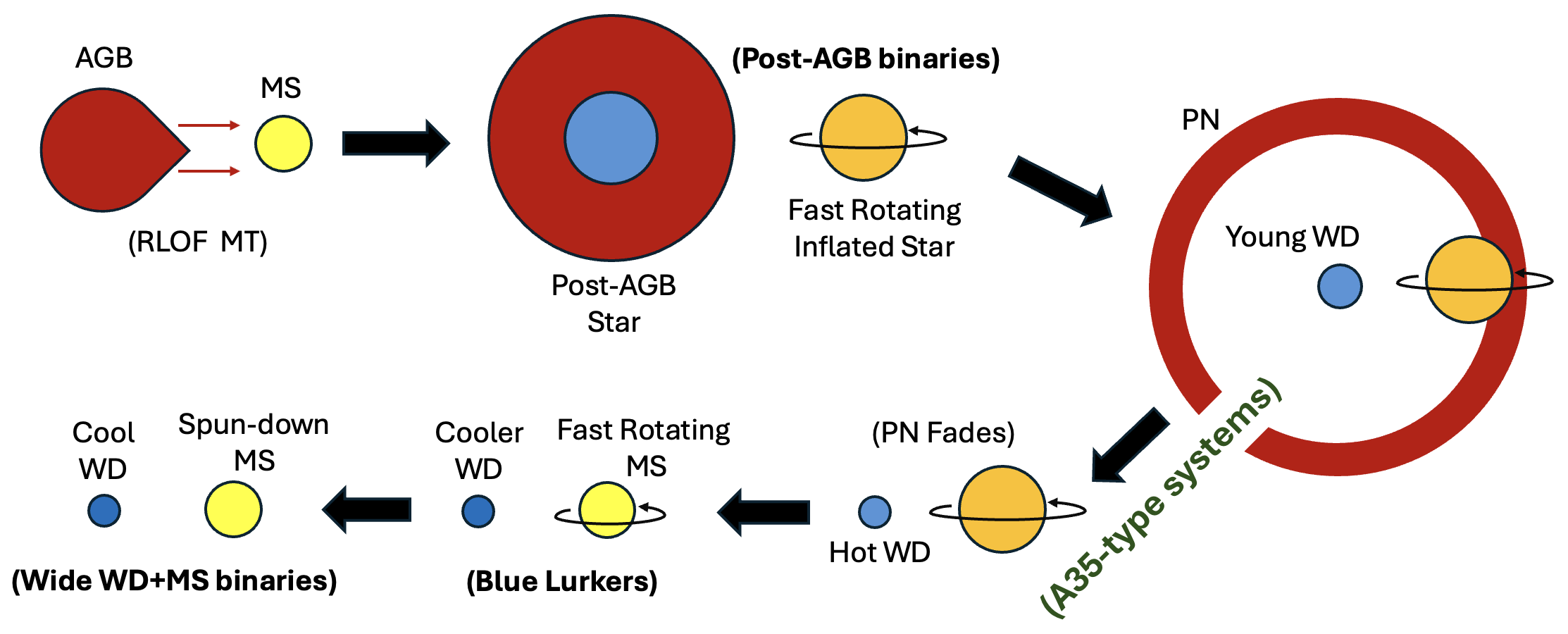}
    \caption{Cartoon depicting the proposed evolutionary sequence connecting the different observed populations of post-mass transfer wide binaries.}
    \label{fig:evolution_cartoon}
\end{figure*}

We now discuss the evolutionary link between A35-type binaries and other post-mass transfer binaries. Figure~\ref{fig:porb_mwd_rappaport} shows A35 in the parameter space of the orbital period, $P_{\rm orb}$, and the WD mass, $M_{\rm WD}$. In the same figure, we plot the analytic $P_{\rm orb}$$-$$M_{\rm WD}$ relation for post-stable mass transfer on the RGB derived in \cite{Rappaport95}. We also show several other classes of post-mass transfer systems namely the Gaia WD+main sequence population \citep{Yamaguchi24b, Shahaf23}, the Kepler SLBs \cite{Kruse14,Kawahara18,Masuda19,Yamaguchi24}, and the UV-excess WD+FGK binaries \citep{Garbutt24}. It is evident that A35 occupies the same position in the parameter space as the other systems. Owing to the lack of reliable masses\footnote{Except for WOCS 14020, but the WD here is proposed to be from a merger product in an initially triple system \citep{Leiner25}. Photometric estimates are available for a few BLs, for those maybe unreliable (note, for example, the significant revision of the mass estimate for WOCS 14020 in \citealt{Nine23} vs \citealt{Leiner25}).} for most Blue Lurkers (BLs, WDs with fast rotating main sequence companions, \citealt{Leiner19,Nine23}), we do not include them in the plot. Future spectroscopic surveys will populate this plot further.

Our MESA models naturally establish a connection between these different populations. We again refer to Figure~\ref{fig:params_vs_age}. As already discussed, A35-type systems constitute the regime where the accretor is thermally relaxing back to main sequence and has just started to spin-down through magnetic braking. Following the models forward, we find that few$\times$$10^{8}$~years later, the white dwarf has cooled to temperatures $\lesssim30$~kK and the accretor has relaxed back to the main sequence. However, it is still rotating rapidly with $P_{\rm rot}\lesssim10$~days. We propose that these constitute the BLs (shaded in red in the figure). Further into the future at several billion years post--mass transfer, the WD has cooled further and the accretor has spun-down significantly. At this stage, the accretors' bulk properties are indistinguishable from those of normal MS stars. This stage constitutes the bulk of wide WD+main sequence stars (primarily from Gaia, shaded in gray in the figure).\footnote{For a preliminary population comparison, we note that \cite{Yamaguchi25} predicts $\sim$$10^4$ WD+MS binaries within $500$~pc. With a median age of a few Gyrs of this population, an accretor inflation timescale of a million years suggests $\sim$$1-5$ A35-type systems within this distance. This is consistent with the known population of these systems. A more thorough population study will be presented elsewhere.} We depict this evolutionary pathway pictorially in Figure~\ref{fig:evolution_cartoon}. We emphasize that this discussion should be interpreted with caution as the specifics of the formation history of each individual system may differ (like the BL WOCS 14020, where the donor is proposed to be a merger product, \citealt{Leiner25}). The unifying scheme, however, is the accretion-induced inflation and subsequent evolution of both the WD and the accretor.


Another related population, which can be viewed as the progenitor population for A35-type binaries, is the post-AGB (or RGB, but we use this term universally for our purpose) binaries \citep{vanWinckel03}. These comprise of a rapidly contracting post-AGB star with a companion (mostly of unknown nature) at orbital periods similar to the other populations \citep{Oomen18}. As the donor has not yet become a WD, these are systems even younger than A35-types with a post-mass transfer age of less than a few thousand years (shaded in brown in Figure~\ref{fig:params_vs_age}). Our models imply that the accretor should be significantly inflated at this stage. The system may thus appear as composed of two evolved stars. In fact, a few such systems are already known, namely HD~172481 \citep{Reyniers01} and SS~Lep \citep{Verhoelst07}. We conjecture that one of the two stars in these binaries is an inflated star. Interestingly, \cite{Jorissen09}, in their Section~2.2, discusses the possibility that several of the stars cataloged as post-AGB can very well be inflated accretors. Our models support this possibility.

\begin{figure}
    \centering
    \includegraphics[width=1\linewidth]{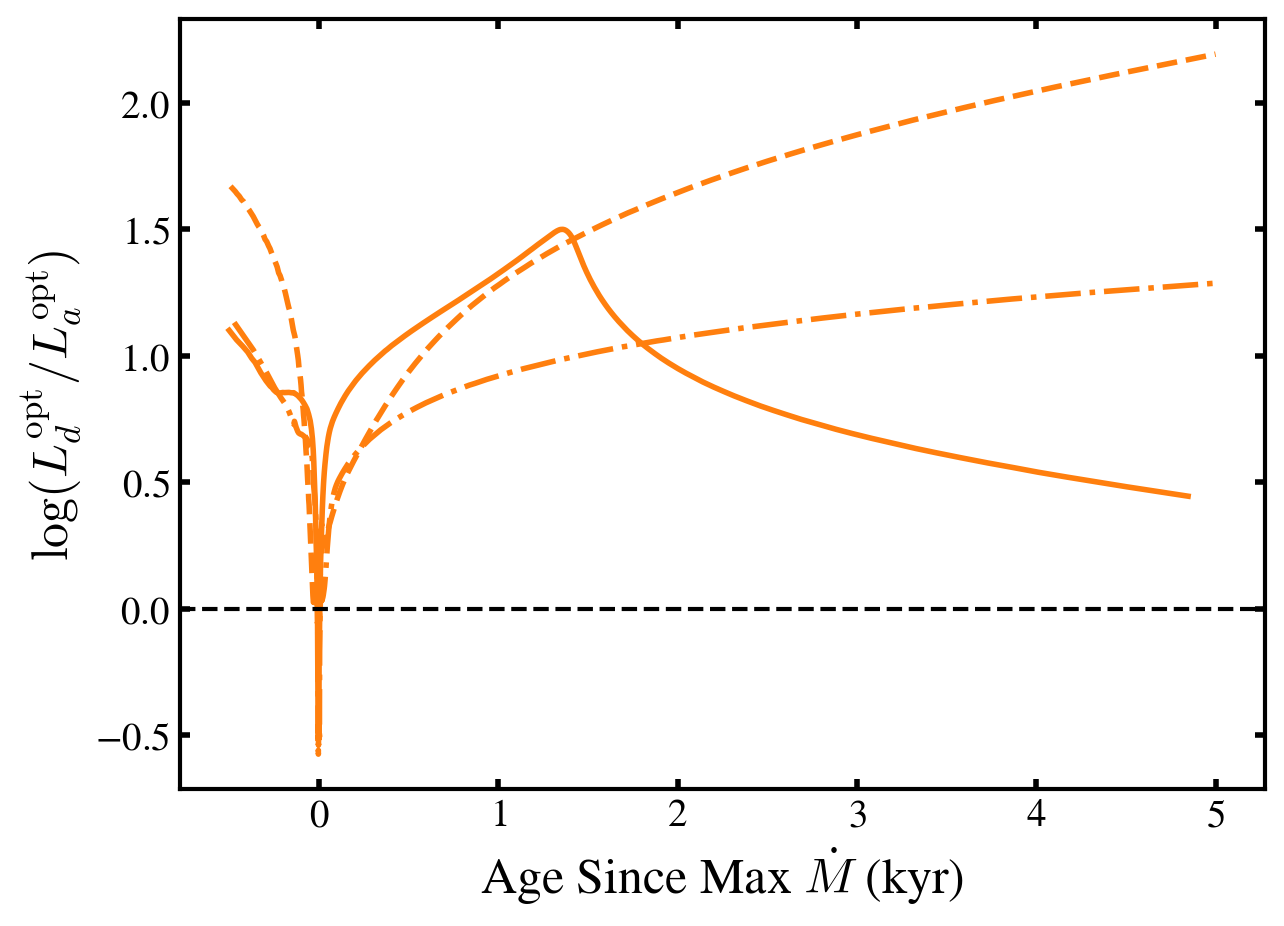}
    \caption{Ratio of the donor to accretor luminosity in the optical ($0.5-0.7$~nm, assuming blackbodies) in the approximate age range of the post-AGB binary systems. In all our models, despite the accretor's inflation, the donor dominates the optical light after peak mass transfer. We follow the same linestyles for the models as in Figure~\ref{fig:radius_vs_temp_a35}.}
    \label{fig:ld_la_optical}
\end{figure}

Most of the companions to post-AGB stars are thought to be main sequence stars \citep{Oomen18}. But most studies arrive at this conclusion by ruling out a WD companion, which does not necessarily rule out giant or subgiant companions. For example, \cite{Hillen16} presents IRAS~08544-4431 as the only post-AGB system where the flux from the companion is detected in the infrared, where they note the possibility of the companion being a red-giant. The other reasoning behind the assumption of a main sequence companion can be the lack of significant flux contribution from the companion in the SED/optical spectra. But, this is not surprising as, despite being inflated, the accretor can be sub-luminous. This is also what we find in our models -- the donor dominates the optical (and bluer) wavelengths till at least $\sim$$10^4$~years post mass transfer (Figure~\ref{fig:ld_la_optical}). The companion can possibly be detected in the infrared (as in \citealt{Hillen16}), which, unfortunately, is often dominated by the dust emission.



\subsection{Observational Predictions}

We now present some predictions which can motivate future observations and test our models:
\begin{itemize}
    \item As the PN timescales are shorter than stellar inflation timescales, there should be more A35-type systems outside PNe. Till now, the search for these systems have mostly been focused inside PNe, significantly limiting the scope of identifying these systems. A broader search for such systems is being undertaken through spectroscopic follow-up of subgiant stars with UV excess, the results of which will be presented elsewhere (Bhattacharjee et al. in prep.).

    \item In all our models, the accretor relaxes back to the main sequence within a few million years of the accretion episode. This time-span is relatively short compared to the WD cooling timescale. Thus, the prediction is that most AU-scale companions to hot (i.e., $T_{\rm eff,\,WD}\gtrsim50$~kK) should appear evolved. In the other way, most AU-scale WD companions to main sequence stars should be `cool'. These are presently in line with 1) with the known WD temperatures in BLs, and 2) lack of any known main sequence companions to hot WDs. This maybe partly influenced by detection bias towards bright giant companions. But given that most of the A35-type systems should have also been detected if the companions were main sequence stars, a significant bias seems unlikely. Nevertheless, more extensive UV spectroscopic follow-up of WD+main sequence binaries (Yamaguchi et al. in prep.) and dedicated searches for companions to hot WDs should be performed.
    


    \item The companions to most post-AGB binaries should appear evolved. Our discussion in the previous section shows that such a possibility cannot yet be ruled out with the current data. Future observations dedicated to the detection and study of the companion are needed.
    
\end{itemize}

\subsection{Caveats}\label{subsec:caveats}

The primary caveat of our models is the uncertainty in the evolution during the highest accretion rates of $\dot{M}$$\gtrsim$$10^{-2}~M_{\odot}\rm/year$. Recent works, for example by \citet{Lu23,Scherbak25}, have shown that there can be mass loss from L2 at such high accretion rates. Though this is most likely to happen when the accretion rate surpasses the Eddington limit -- not the case with our models -- we do not rule an L2 outflow. Also, the donors in our model often fill the outer lobe (Figure~\ref{fig:stability_analysis_rlof_olof}), which lead to L3 mass loss. Both these would have the same effect of draining a significant amount of angular momentum, shrinking the orbit and likely triggering CE. In MESA, L2 mass loss can in principle be mimicked using the $\delta$ mass loss parameter, which drains angular momentum through a circumbinary disk. But we do not attempt this as such models tend to be highly unstable and uncertain. Additionally, owing to the large Roche lobe overfill factors in our models (Figure~\ref{fig:stability_analysis_rlof_olof}), the epochs of highest $\dot{M}$, including the maximum inflated radius, should be interpreted with caution. However, the evolution of the accretor post-mass transfer is expected to be sufficiently reliable that our proposed evolutionary picture is not affected by these limitations.



A second caveat concerns the effects of rotation and magnetic fields, which we have not self-consistently modeled in the MESA calculations. Both of these can provide additional pressure support, which can assist in the inflation of the star (see Appendix~\ref{app:mesa_with_rotation}). A spin-up may also enhance stellar winds which can drain angular momentum in a more complicated way than assumed in the magnetic braking law incorporated in our calculations. Additionally, a 1D evolution code like MESA is not enough to capture all the complicated physics. In the future, detailed 3D simulations of interacting binaries may shine further light on how stars respond to rapid accretion

\section{Summary and Conclusions}\label{sec:conclusions}

Abell~35 (A35) type systems are peculiar binaries comprised of a very hot and young WD (with effective temperatures of $\sim$$10^5$~K) and a fast-rotating subgiant companion in a `wide' ($\sim$$10^3$~day period) orbit. There are eight known systems, six of which are young enough to be inside planetary nebulae. The fast rotation of the subgiant hints at a recent accretion episode, presumably from the WD progenitor giant star. In this work we revisit this population of systems, primarily to shed light on their formation channel and establish their significance in the broader context of post-mass transfer evolution. Specifically, we argue that the subgiants in these systems are likely to be main sequence stars temporarily inflated through the recent accretion episode. We anchor our analysis to the prototype system A35 -- the only system to have an astrometric binary solution in Gaia DR3. Firstly, the Gaia parallax yields a distance of $165.75$~pc, significantly closer than assumed in \cite{Jacoby81} and \cite{Ziegler12}. The system has an orbital period of $790~\rm d$, comparable to other populations of au-scale WD+MS binaries recently identified with Gaia. In light of this astrometric orbit and aided by new spectroscopy of the subgaint companion, we construct MESA binary evolution models to study the inflation of the accretor. Below, we present our inferences in further detail.


Our primary inferences from the joint analysis of Gaia astrometry, SED and high-resolution spectrum of A35 are as follows:

\begin{enumerate}
    \item The subgiant parameters: A fit of the spectral energy distribution (SED) yields a temperature of $4925\pm75$~K and a radius of $2.95\pm0.05~R_{\odot}$ (Figure~\ref{fig:a35_sed_fit}). Analysis of the FEROS spectrum yields a high surface rotation speed of $v\sin(i)=86\pm5~\rm km~s^{-1}$, which corresponds to a surface rotation speed of $v_{\rm rot}\approx196\rm km\,s^{-1}$. This is near the critical rotation velocity and substantially higher than found in normal subgiants. Comparing the optical spectrum of the subgiant in A35 with synthetic templates and stellar libraries, we arrive at a metallicity of $\rm[Fe/H]$$\approx$$0.0\pm0.15$ (Figure~\ref{fig:spectrum_comparison}). Unlike \cite{Thevenin97}, we do not find any significant barium enhancement.

    \item The white dwarf parameters: Combining astrometry and UV spectroscopy, we find the WD radius to lie in the range of $\approx$$(1.75\pm0.25)\times10^{-2}~R_{\odot}$. We show this to be in tension with the spectroscopic $\log(g)=7.2\pm0.3$ and temperature ($80\pm10$~kK) estimated in \cite{Ziegler12}, as there is no consistent solution as per the post-AGB models of \cite{Bertolami16}. Allowing for higher $\log(g)$ (as in \citealt{Herald02}) can yield a possible solution at $M_{\rm WD}\gtrsim0.57~M_{\odot}$ (Figure~\ref{fig:a35_millerb_compare_a35_mist_compare}, left panel).

    \item Comparison of the Gaia astrometry-allowed subgiant masses to a range of initial masses of the WD shows that it is unlikely that the subgiant is a naturally evolved star (Figure~\ref{fig:minit_msbg_compare}). This favors the scenario of the subgiant being a temporarily inflated main sequence star.
    

    \item Gaia astrometry shows that A35 is the inner binary of a hierarchical triple system, with the tertiary being a K dwarf at a distance of $\approx$$1600$~AU. Though its influence is negligible today, it may have played a significant role in the past in bringing the inner binary to its current configuration and triggering mass transfer. 

\end{enumerate}

Our primary results from the MESA binary models are as follows:

\begin{enumerate}

    \item We use a new accretion prescription where we use realistic relations for the energy, density, and pressure of the in-falling material (Equation~\ref{eq:acc_precription}). This is expected to capture the entropy deposited on the accretor better than the fiducial MESA setup, especially at high accretion rates. This prescription allowed us to evolve both stars in MESA binary simulations self-consistently through RGB/AGB mass transfer phases even for high initial donor to accretor mass ratios.
    

    \item The accretor inflates significantly during mass transfer. For example, a $0.5~M_{\odot}$ accretor achieves a peak radius of $\approx$$80~R_{\odot}$ during the phase of highest $\dot{M}$ (Figure~\ref{fig:radius_vs_temp_a35_model_detail}). However, it doesn't fill its Roche lobe, contrary to what is commonly assumed for inflated companions. Post--mass transfer, the inflated accretor begins to thermally relax back to main sequence. It remains significantly inflated for at least the next million years (i.e., comparable to its thermal timescale). 

    \item The phase of thermal relaxation meets the parameters of the A35-type systems reasonably well at a post-mass transfer age of $10^4-10^6$~years (Figures~\ref{fig:radius_vs_temp_a35} and \ref{fig:params_vs_age}). Additionally, we present other post--mass transfer systems, namely the post-AGB binaries, the Blue Lurkers, and the broader population of WD+main sequence binaries as representing different stages of the same evolutionary pathway (Figure~\ref{fig:evolution_cartoon}). 
    
    \item The primary prediction from our models is that most AU-scale companions to hot (cold) WDs should be inflated (normal main sequence) stars. This can be tested through optical and UV follow-ups of binaries with a WD, usually selected through UV excess or being located inside PNe.
\end{enumerate}



\section{Acknowledgments}

We thank the anonymous referee for their very valuable comments. S.B. thanks David Jones for reading the manuscript and providing comments. This research was supported by NSF grants AST-2508988 and AST-2307232. S.B. thanks Osmar Rodriguez Suarez for pointing out the error underestimation in the reddening estimate. C.S. acknowledges support from the Department of Energy Computational Science Graduate Fellowship, supported by the U.S. Department of Energy, Office of Science, Office of Advanced Scientific Computing Research, under Award Number DE-SC0026073. We have used the \texttt{Python} packages Numpy \citep{harris2020array}, SciPy \citep{2020SciPy-NMeth}, Matplotlib \citep{Hunter:2007}, Pandas \citep{reback2020pandas}, Astropy \citep{Astropy13, Astropy18}, and Astroquery \citep{astroquery19} at various stages of this research.

\section{Data Availability}

All the relevant codes to reproduce the results of this paper can be found in \url{https://github.com/Soumin1908/InflatedStars_A35_MesaModels.git}

\appendix

\section{The possible dynamical role of the tertiary}\label{subsec:triple}

Hierarchical triples are common in the Galactic field \citep[e.g.,][]{Raghavan2010,Tokovinin14a,Moe17,Winters19,MoeKratter21,Offner23,Shariat25_10k} and play a key an important role in the formation and evolution of post--mass-transfer WD+main sequence binaries such as A35 \citep[e.g.,][]{Toonen2016,Toonen20,Toonen2022,Knigge22,Rajamuthukumar23,Rajamuthukumar25,Hallakoun24,Shariat23,Shariat25_CV,Shariat25_Merge, Shariat26_FRB, Perets25}. In particular, if the inner binary formed on a relatively wide orbit ($\gtrsim10$ au), secular perturbations from the tertiary would have driven oscillations to the inner binary's eccentricity and inclination through the eccentric Kozai-Lidov mechanism \citep{Kozai62,Lidov62,Naoz16}. At fixed semi-major axis, these eccentricity oscillations amount to oscillations in the periastron distance of the inner binary. As the primary evolves onto the giant branch and its radius increases, close periastron passages allow tides to efficiently dissipate orbital energy, causing the orbit to circularize and shrink. Since the tidal synchronization timescale for equilibrium tides depends steeply on stellar radius \citep[$\propto R^6$;][]{Zahn77}, the most likely mass-transfer outcome is that the inner-binary interaction begins when the primary is on the late RGB or AGB \citep[e.g.,][]{Toonen20,Shariat25_CV, Shariat26_FRB}. At this stage, the tertiary’s gravitational influence is negligible, and the inner binary evolves effectively in isolation, allowing stable mass transfer and the subsequent binary evolution to proceed.

Overall, the primary role of hierarchical triples in A35-like systems is to increase the fraction of binaries that experience mass transfer by enabling close periastron passages in binaries that would otherwise remain too wide ($\gtrsim10$ au) to interact in isolation. The detailed binary evolution calculations presented in Section~\ref{sec:mesa_modeling} are agnostic to the pre–mass-transfer history of the system. Nevertheless, tertiary companions may be common and astrophysically important to the formation of A35-like systems today. Note that some tertiaries may no longer be bound due to sudden mass loss either during mass transfer or late-AGB mass loss in the inner binary \citep[e.g.,][]{ElBadry18,Igoshev20,Shariat23,Hwang25,OConnor25}, while others may have evolved into faint white dwarfs that are challenging to detect observationally. Future systematic searches for tertiaries around post--mass-transfer WD+main sequence binaries, particularity in PNe, will be essential for quantifying their occurrence, separations, and evolutionary role.

\begin{figure*}
    \centering
    \includegraphics[width=1\linewidth]{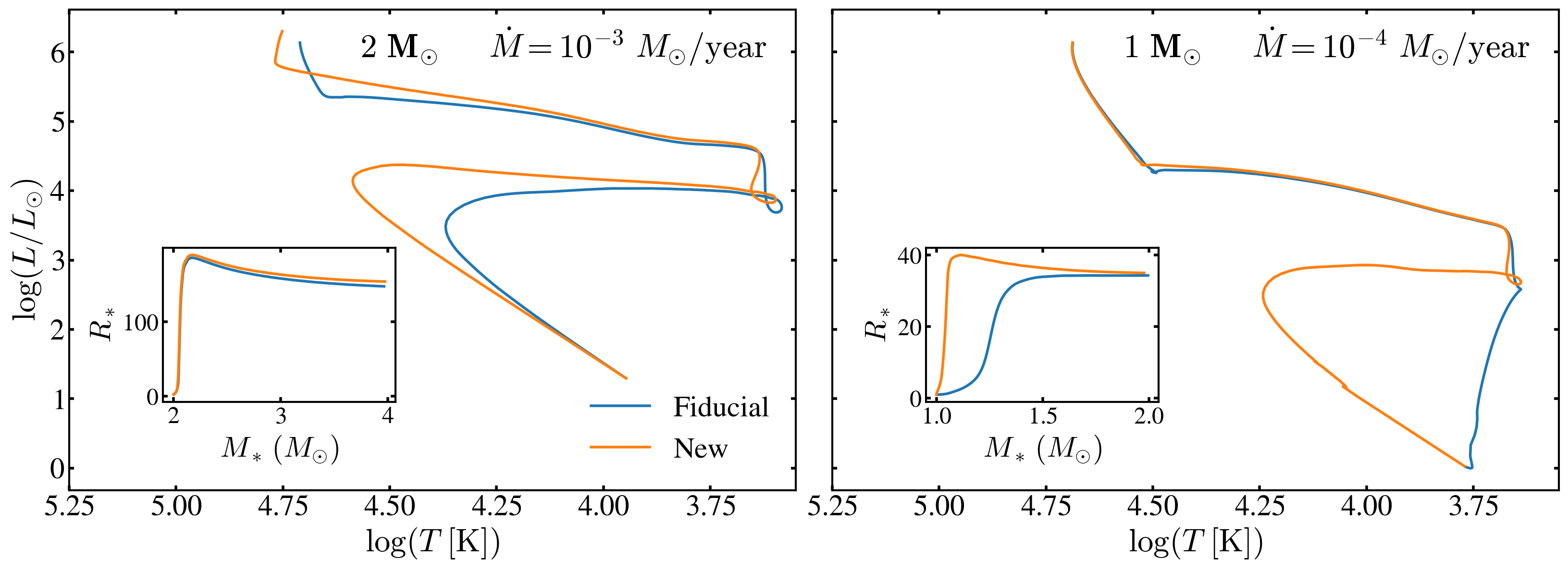}
    \caption{Evolution of a $2~M_{\odot}$ accretor with $\dot{M}=10^{-3}~M_{\odot}\rm/year$ (left panel) and $1~M_{\odot}$ accretor with $\dot{M}=10^{-4}~M_{\odot}\rm/year$ (right panel) for the fiducial and new accretion prescriptions. The insets show the evolution of the accretor radius until its mass doubles. Unlike the fiducial case, the added entropy in the new accretion prescription disrupts the convective envelopes of low-mass stars at early stages, thus mimicking the evolution of high-mass stars.}
    \label{fig:comparison_with_hamstar}
\end{figure*}

\section{Companion Mass from Gaia Astrometry}\label{app:comp_mass_kepler}

The six astrometric parameters provided by Gaia are relevant for this calculation are: the parallax, $\varpi$, the orbital period, $P_{\rm orb}$, and the four Thiele–Innes (TI) elements $A$, $B$, $F$, and $G_T$ (to distinguish from the gravitational constant). We define $P=A^2+B^2+F^2+G_T^2$ and $Q = AG_T - BF$. From the definition of the TI elements, it is then easy to show that the orbital inclination is given by:
\begin{equation}
    \cos(i)=\frac{P-\sqrt{P^2-4Q^2}}{2Q},
\end{equation}
which yields a corresponding semi-major axis of the Gaia photo-center:
\begin{equation}
    a_{\rm pc}=\sqrt{\frac{Q}{\cos(i)}}\frac{1}{\varpi}.
\end{equation}
With a WD to subgiant Gaia G-band flux ratio of $\epsilon=F_{G,~\rm WD}/F_{G,~\rm SG}$, this yields a binary semi-major axis as:
\begin{equation}
    a = a_{\rm pc}\left(\frac{M_{\rm WD}}{M_{\rm SG}+M_{\rm WD}}-\frac{\epsilon}{1+\epsilon}\right)^{-1}.
\end{equation}
We solve this consistently with Kepler's third law to calculate $M_{\rm SG}$ for a given $M_{\rm WD}$.

\section{Comparison with the `Hamstars'}\label{app:comparison_with_hamstars}

\begin{figure}
    \centering
    \includegraphics[width=1\linewidth]{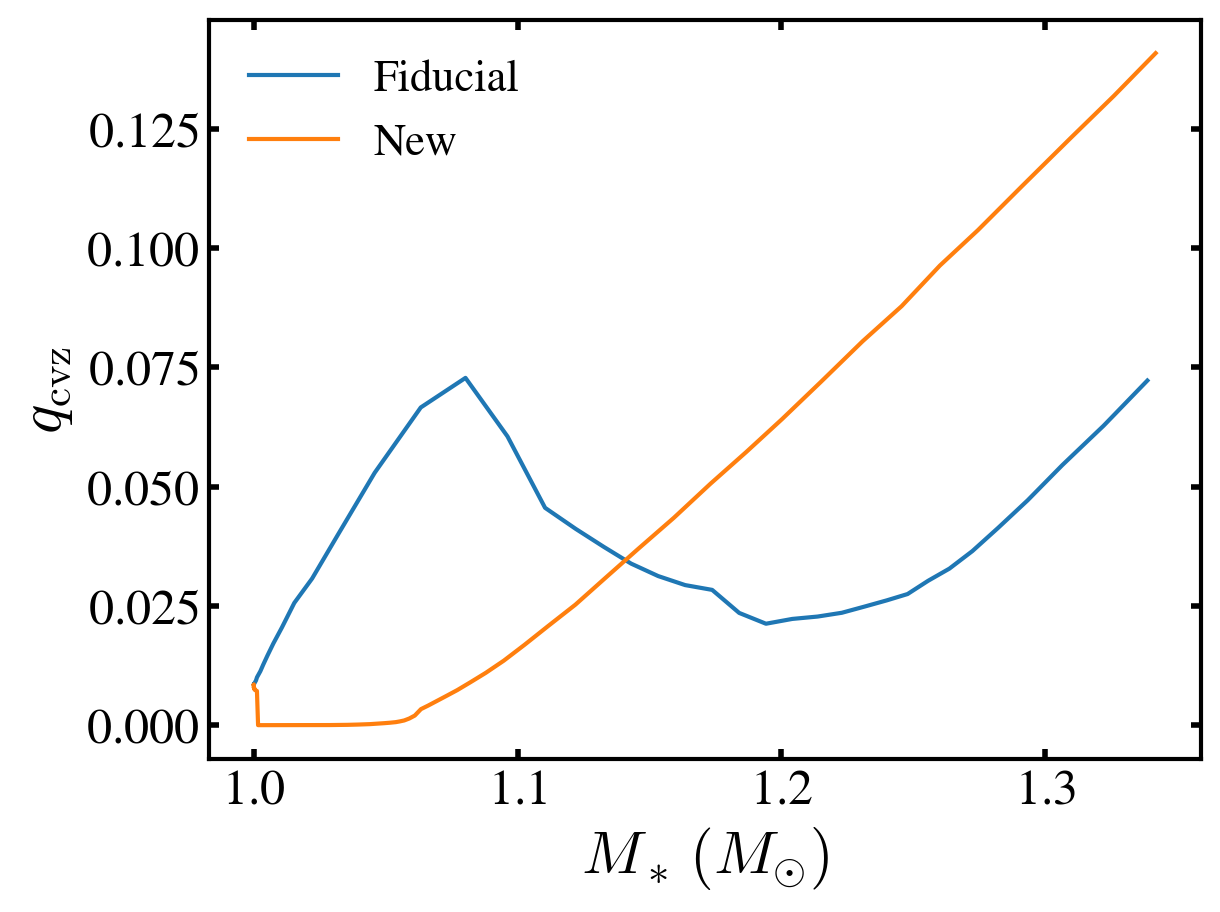}
    \caption{Mass fraction of the surface convection zone (calculated from the Ledoux unstable mass at radius $>$$0.5~R_{\odot}$) as a function of the total mass for the $1~M_{\odot}$ accretor for both accretion prescriptions. The new prescription disrupts surface convection almost completely at the very initial stages of accretion.}
    \label{fig:qcvz_vs_mass}
\end{figure}

Recently \cite{Lau24} conducted a systematic study of inflation of accreting stars of masses $>$$2~M_{\odot}$. They assumed a constant accretion rate, and continued accretion until the total mass of the accretor becomes $100~M_{\odot}$. They also employed the fiducial \texttt{MESA} accretion prescription of the accreting material being of the same physical properties of the stellar photosphere. They call their systems `Hamstars' for their systems are objects which swallow mass and inflate.

Apart from the difference in the accretion prescription used, the other major difference with this work is the range of stellar masses of interest. The mass range investigated in \cite{Lau24} have radiative envelopes. Whereas, this works explores the mass $\lesssim$$1~M_{\odot}$, which has a convective envelope. This can potentially lead to a substantial difference in their reaction to accretion. This is because, convective polytropes tend to prevent the expansion of the star, whereas radiative envelopes tend to expand on accretion. This was recently investigated by \cite{Zhao24}.

We discuss here how our new accretion prescription affect the response of the accretor. We do this by taking a $1~M_{\odot}$ and $2~M_{\odot}$ star and conducting the exercise as in \cite{Lau24}: adopt a constant $\dot{M}$ for a long duration. For demonstration, we choose $\dot{M}=10^{-4}~M_{\odot}\rm/year$ and $\dot{M}=10^{-3}~M_{\odot}\rm/year$ respectively for the two stellar masses. We ramp up the accretion rate to the adopted value over a duration such that a net of $0.1~M_{\odot}$ is accreted during this phase. We checked that our results are not affected with the change in the ramp-up duration.

The results are shown in Figure~\ref{fig:comparison_with_hamstar}, where we present the evolution of the accretor in the Hertzsprung-Russel diagram (HRD) and the change in stellar radius as a function of mass (till the mass is doubled) as insets. We first discuss the case of the $2~M_{\odot}$ star which has a radiative envelope. For the fiducial prescription of MESA (thermodynamic properties of the accreted material being the same as stellar surface), we recover the tracks presented in \cite{Lau24}. But with the new prescription, we see that the star becomes much brighter compared to the fiducial case before approaching the Hayashi limit, after which the evolution is nearly identical. This is primarily caused by a higher temperature which results from the higher accretion energy in the new prescription. However, the radius evolution stays nearly identical throughout. Why this is the case is unclear.

The case of the $1~M_{\odot}$ star, however, is very different -- the HRD tracks for the two prescriptions differ significantly. This is accompanied by a substantial difference in radius evolution, where we see that with the new prescription, the star undergoes a rapid expansion at the very early stages of the mass transfer. Interestingly, this behavior is qualitatively similar to that of the $2~M_{\odot}$ star. Inspection shows that the envelope properties of the low mass star evolves differently with the two prescriptions (Figure~\ref{fig:qcvz_vs_mass}). In the fiducial case, the added mass bears the same entropy as the stellar surface. Thus, the envelope remains convective. This makes the star evolve similar to a convective star, and it almost follows the Hayashi track from the beginning. With the new prescription, however, the entropy of the added material is much higher than that of the stellar surface. This quickly disrupts the surface convection, making the star behave qualitatively similarly to a radiative-envelope star. Once it reaches the Hayashi limit, the two tracks converge and the subsequent evolution is nearly identical.

This exercise demonstrates the importance of properly considering the accretion entropy deposited on the stellar evolution, especially for low mass stars. We deem this particularly important for close mass transfer binaries, where a quick initial expansion of the accretor can result in an early CE. This motivates future work to quantify the effect more accurately.

\section{Effects of Accretor Rotation}\label{app:mesa_with_rotation}

\begin{figure}
    \centering
    \includegraphics[width=1\linewidth]{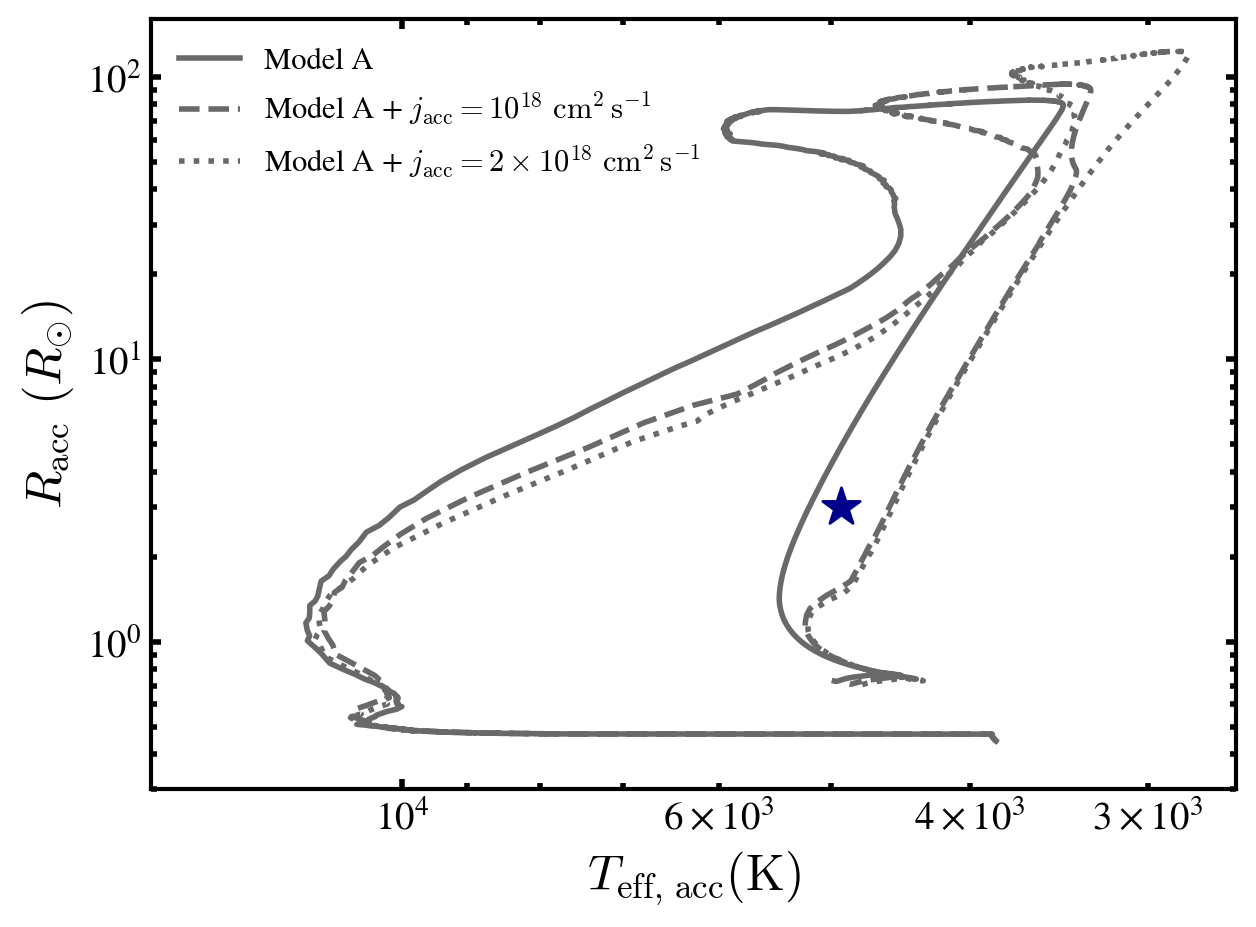}
    \caption{Comparison of the accretor evolution in the $R-T_{\rm eff}$ space (same as top panel of Figure~\ref{fig:radius_vs_temp_a35}) between the non rotating (solid line) and rotating (dashed and dotted lines) models. The specific angular momentum of the accreting material used the rotating models are mentioned in the legend. The rotating models are cooler and inflates to larger radii (still remaining within Roche lobe), but otherwise exhibits similar evolution to its non-rotating counterpart.}
    \label{fig:radius_vs_temp_a35_withrot}
\end{figure}

\begin{figure}
    \centering
    \includegraphics[width=1.0\linewidth]{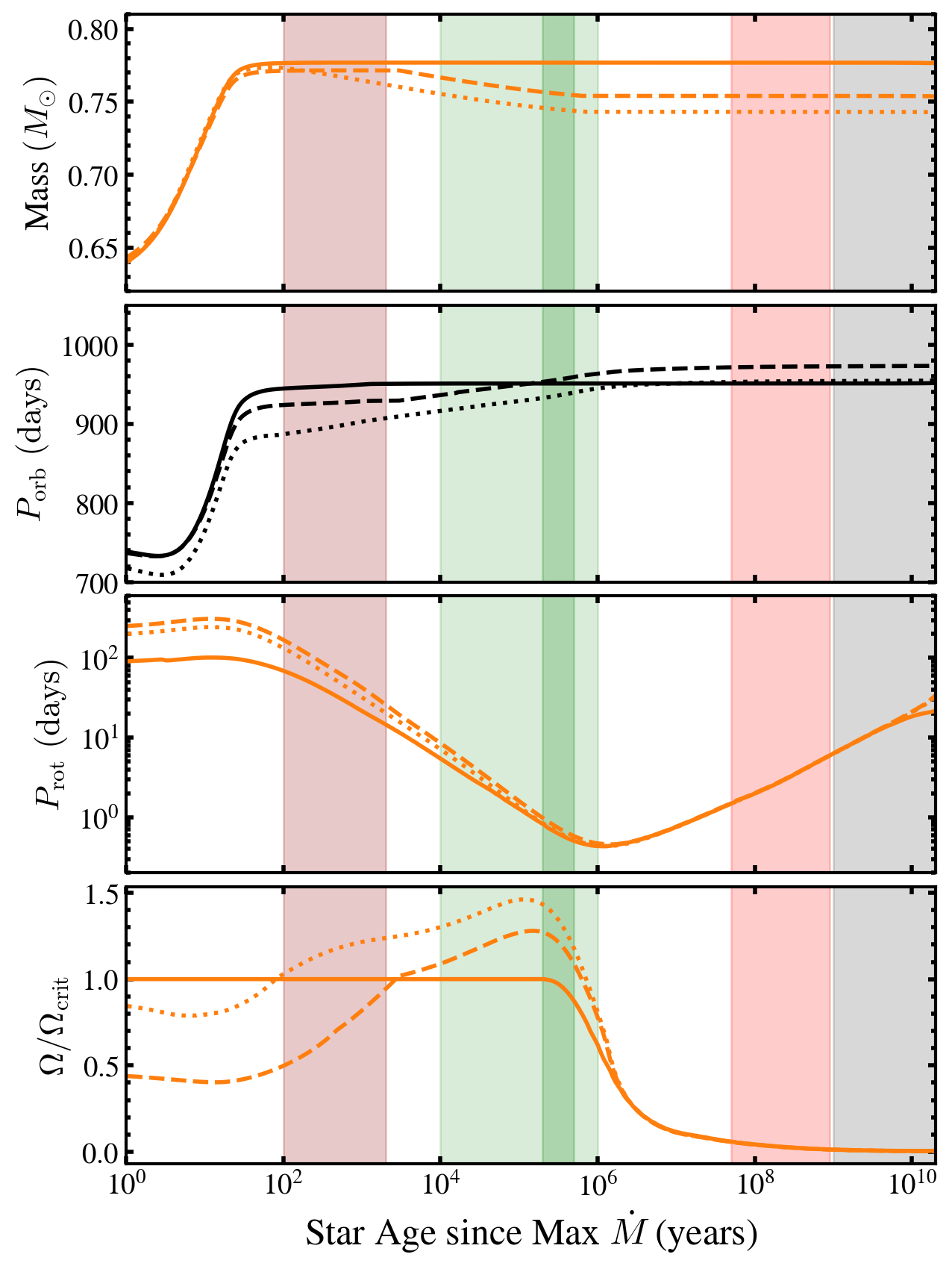}
    \caption{Evolution of the accretor's mass, binary period, accretor's rotation period and $\Omega/\Omega_{\rm crit}$ (top to bottom) for the non-rotating and rotating models. The linestyle followed is the same as Figure~\ref{fig:radius_vs_temp_a35_withrot}. The rotating models yields lower final accretor mass and longer orbital period.}
    \label{fig:params_vs_age_all_models_withrot}
\end{figure}

We demonstrate here the possible effects of accretor rotation in MESA. We turn on accretor rotation by setting \texttt{new\_rotation\_flag = .true.} and \texttt{change\_rotation\_flag = .true.} in our inlist. The standard prescription for momentum gain at the accretor's surface Keplerian velocity is enabled by setting \texttt{do\_j\_accretion = .true.}, but this leads to unphysically large rotation rates for the accretor. Owing to the high accretion rates, the enhanced wind fails to keep up in draining enough angular momentum from the star. Thus the accretor attains $\Omega/\Omega_{\rm crit}$ values (given by the variable \texttt{surf\_avg\_omega\_div\_omega\_crit}) of substantially larger than unity at the very early stages of mass transfer and it remains so for the remainder of the simulation. These models are, thus, likely highly unphysical and we do not show them.

Instead, for the purpose of this demonstration, we thus manually set the specific angular momentum of the accreting material to a constant value using \texttt{accreted\_material\_j}. We choose values such that the accretor spins up significantly and the issue is mitigated, though not completely solved -- $\Omega/\Omega_{\rm crit}$ still overshoots unity, but by a much smaller margin and for a shorter duration. We present the results only for the Model~A initial conditions (see Section~\ref{subsubsec:model_overview}), but the effects on the other models are qualitatively the same. To implement magnetic braking in the accretor, we set \texttt{use\_other\_torque = .true.} and implemented Equation~\ref{eq:mb} in \texttt{run\_star\_extras}. We note that the purpose of these rotating calculations is not to construct a self-consistent theory of angular momentum transport during accretion but to demonstrate that the bloated accretor scenario remains viable if the star is allowed to spin up to values comparable to those observed. The amount of angular momentum ultimately retained by the accretor is uncertain because it depends on poorly understood transport processes within the boundary layer, the accretion disk, magnetic torques, and outflows \citep{Rafikov16,Arzamasskiy18,Cohen26}.

We present the new radius-temperature evolution of the accretor in Figure~\ref{fig:radius_vs_temp_a35_withrot}, compared to the evolution without rotation presented in Section~\ref{subsec:mesa_results}. We find that the evolutionary track shifts towards lower $T_{\rm eff}$. We interpret this to be analogous to gravity darkening -- the enhanced centrifugal force sustains hydrostatic equilibrium at a larger radius and lower $T_{\rm eff}$ for the same luminosity. In fact, we see that the models with rotation is cooler than A35 during the thermal contraction phase. This echoes the discussion in Section~\ref{subsubsec:a35_compare} that stellar rotation can assist in meeting the parameters of the inflated star in A35-type systems better. In addition, the accretor inflates to larger radii with rotation. The maximum radius is $\approx$$14\%$ and $\approx$$50\%$ larger in the slower and faster rotating models, respectively, compared to the non-rotating model. In the latter model, the accretor fills nearly $80\%$ of its Roche lobe at its maximum radius. Thus, we acknowledge that it is possible for rotation-assisted expansion of the accretor to fill its Roche lobe and trigger CEE. This demands a more detailed analysis of accretor responses in future works.

The evolution of accretor properties and binary period are presented in Figure~\ref{fig:params_vs_age_all_models_withrot}. We first note that $\Omega/\Omega_{\rm crit}$ reaches $\approx$$1.3-1.5$ during the thermal contraction phase. This is unphysical, but mitigating this requires more sophisticated treatment of winds and angular momentum loss which is beyond the scope of this paper. Owing to the high rotation during this contraction phase, the accretor loses non-negligible amount of mass through enhanced winds, making the final accretor mass slightly smaller and the orbital period slightly longer than the non-rotating models.
The long-term spin period evolution is set by magnetic braking and is almost identical in the two models. 

Overall, the effect of rotation on the accretor's observables can be summarized as: 1) lowering $T_{\rm eff}$ for a given radius, 2) enhanced mass loss, especially during the contraction phase, and 3) mildly longer orbital periods compared to the non rotating model. These effects are small but important for quantitative comparison with observations. For example, though the radius and $T_{\rm eff}$ changes in the required direction for A35, the final orbital period moves away from that observed. Qualitatively, however, our conclusions remain unaffected.


\bibliography{ref}{}
\bibliographystyle{aasjournal}



\end{document}